\documentclass[twocolumn,trackchanges]{aastex62}
\usepackage{hyperref}
\usepackage{amsmath}
\usepackage{color}
\usepackage{bm}

\shorttitle{Heavy Element Absorption and Ionization at High Redshift}
\shortauthors{Cooper et al.}

\providecommand{\civ}{\ion{C}{4}}
\providecommand{\mgii}{\ion{Mg}{2}}
\providecommand{\feii}{\ion{Fe}{2}}
\providecommand{\hi}{\ion{H}{1}}
\providecommand{\oi}{\ion{O}{1}}
\providecommand{\ovi}{\ion{O}{6}}
\providecommand{\cii}{\ion{C}{2}}
\providecommand{\ciii}{\ion{C}{3}}
\providecommand{\siii}{\ion{Si}{2}}
\providecommand{\alii}{\ion{Al}{2}}
\providecommand{\siiv}{\ion{Si}{4}}

\providecommand{\heii}{\ion{He}{2}}
\providecommand{\lya}{Ly$\alpha$}
\providecommand{\nhi}{\ensuremath{N_{\rm HI}}}

\providecommand{\xpm}[3]{\ensuremath{#1^{+#2}_{-#3}}}
\providecommand{\logN}[1]{\ensuremath{\log N_{\text{#1}}}}
\providecommand{\bD}[1]{\ensuremath{b_{\text{#1}}}}
\providecommand{\X}[1]{\ensuremath{N_{\text{#1II}}}/N_{\text{#1IV}}}
\providecommand{\kms}{\text{km s}\ensuremath{^{-1}}}
\providecommand{\N}[1]{\ensuremath{N_{\text{#1}}}}
\newcommand{\dNdX}{\ensuremath{\frac{d\mathcal{N}}{dX}}}
\newcommand{\dNdXt}{\ensuremath{d\mathcal{N}/dX}}

\begin{document}

\title{Heavy Element Absorption Systems at $5.0<z<6.8$:
\\Metal-Poor Neutral Gas and a Diminishing Signature of Highly Ionized Circumgalactic Matter}

\author{Thomas J. Cooper}
\affiliation{MIT-Kavli Center for Astrophysics and Space Research, 77 Massachusetts Avenue, Cambridge, MA 02139, USA}
\affiliation{The Observatories of the Carnegie Institution of Science, 813 Santa Barbara Street, Pasadena, CA 91101, USA}
\author{Robert A. Simcoe}
\affiliation{MIT-Kavli Center for Astrophysics and Space Research, 77 Massachusetts Avenue, Cambridge, MA 02139, USA}
\author{Kathy L. Cooksey}
\affiliation{University of Hawai`i at Hilo, 200 West K\=awili Street, Hilo, HI 96720, USA}
\author{Rongmon Bordoloi}
\affiliation{MIT-Kavli Center for Astrophysics and Space Research, 77 Massachusetts Avenue, Cambridge, MA 02139, USA}
\affiliation{Hubble Fellow}
\affiliation{Department of Physics, North Carolina State University, Raleigh, NC 27695, USA}
\author{Daniel R. Miller}
\affiliation{MIT-Kavli Center for Astrophysics and Space Research, 77 Massachusetts Avenue, Cambridge, MA 02139, USA}
\author{Gabor Furesz}
\affiliation{MIT-Kavli Center for Astrophysics and Space Research, 77 Massachusetts Avenue, Cambridge, MA 02139, USA}
\author{Monica L. Turner}
\affiliation{MIT-Kavli Center for Astrophysics and Space Research, 77 Massachusetts Avenue, Cambridge, MA 02139, USA}
\affiliation{Las Cumbres Observatory, 6740 Cortona Drive, Suite 102, Goleta, CA 93117-5575, USA}
\author{Eduardo Ba\~nados}
\affiliation{The Observatories of the Carnegie Institution of Science, 813 Santa Barbara Street, Pasadena, CA 91101, USA}

\correspondingauthor{Thomas Cooper}
\email{tcooper@carnegiescience.edu}


\begin{abstract}

\noindent Ratios of different ions of the same element encode ionization information independently from relative abundances in quasar absorption line systems, crucial for understanding the multiphase nature and origin of absorbing gas, particularly at $z>6$ where \hi\ cannot be observed. Observational considerations have limited such studies to a small number of sightlines, with most surveys at $z>6$ focused upon the statistical properties of individual ions such as \mgii\ or \civ. Here we compare high- and low-ionization absorption within 69 intervening systems at $z>5$, including 16 systems at $z>6$, from Magellan/FIRE spectra of 47 quasars together with a Keck/HIRES spectrum of the `ultraluminous' $z=6.3$ quasar SDSSJ010013.02+280225.8. The highest redshift absorbers increasingly exhibit low-ionization species alone, consistent with previous single-ion surveys that show the frequency of \mgii\ is unchanging with redshift while \civ\ absorption drops markedly toward $z=6$. We detect no \civ\ or \siiv\ in half of all metal-line absorbers at $z>5.7$, with stacks not revealing any slightly weaker \civ\ just below our detection threshold, and most of the other half have $\N{CII}>\N{CIV}$. In contrast, only 20\% of absorbers at 5.0--5.7 lack high-ionization gas, and a search of 25 HIRES sightlines at $z\sim3$ yielded zero such examples. We infer these low-ionization high-redshift absorption systems may be analogous to metal-poor Damped Lyman-$\alpha$ systems ($\sim1\%$ of the absorber population at $z\sim3$), based on incidence rates and absolute and relative column densities. Simple photoionization models suggest that circumgalactic matter at redshift six has systematically lower chemical abundances and experiences a softer ionizing background relative to redshift three.

\end{abstract}

\section{Introduction}

The ionization and chemical abundance of intergalactic and circumgalactic gas should evolve with redshift, as the metagalactic background radiation spectrum changes and chemical enrichment occurs. In principle, quasar absorption lines are sensitive to these effects, but in practice the evolution only manifests in subtle ways, and observational limitations that change with redshift introduce additional obstacles.

For example, early measurements of the \civ\ to \siiv\ ratio revealed tentative evidence of heating from \heii\ reionization at $z\sim3$ \citep{1998AJ....115.2184S}. However this result was not confirmed using other samples at similar redshift \citep{2002A&A...383..747K,2011ApJ...729...87C,2015ApJS..218....7B}, and yet it may be detected at higher redshift using measurements with X-Shooter \citep{2013MNRAS.435.1198D}.

Likewise, \citet{2013ApJ...764....9M} found that carbon, silicon, iron and aluminum ions measured in \mgii-selected absorbers have nearly indistinguishable equivalent width distributions---in both absolute value and relative ratios---at all redshifts probed between $0<z<5$. Below $z\sim 5$ one can measure $N_{\rm HI}$, and an increasing proportion of \mgii ~systems are associated with neutral Damped \lya\ absorption (DLA; $N_{\rm HI}>10^{20.3}$ cm$^{-2}$) and slightly ionized sub-DLAs toward higher redshift. Indeed by $z\sim 3.5$ all \mgii\ with measured \nhi\ in their sample show $\N{HI}>10^{19} \mathrm{~cm}^{-2}$, compared to just 15\% at $z\sim 1$.

As reionization is approached at $z>6$, it becomes impossible to identify discrete \hi\ absorbers yet the incidence rate of low-ionization heavy element lines remains robust. Broadly defined, the low-ionization species (e.g. \mgii, \oi, \cii, \siii, \feii, \alii) represent states from which valence electrons can be ionized by photons with $1<E_{\rm ion}<2$ Ryd. Their presence at $z>6$ was first remarked after the detection of numerous \oi ~lines in the spectrum of a single bright high-redshift QSO \citep{2006ApJ...640...69B}, and then more systematically studied via \oi\ $\lambda 1302$\AA, \cii\ $\lambda 1334$\AA, and \siii\ $\lambda 1260$\AA, whose transitions are close in rest wavelength to \lya\ and therefore accessible in optical spectra \citep{2011ApJ...735...93B}. Careful statistical surveys for low-ionization \mgii\ doublets in infrared quasar spectra \citep{2017ApJ...850..188C,2012ApJ...761..112M, 2017MNRAS.470.1919B} found that for equivalent widths $W_{r,2796}<1.0$\AA---i.e. the majority of the population---there is no evidence for evolution in the comoving incidence rate, out to the highest redshifts probed at $6<z<7$ (though stronger \mgii\ systems do appear to peak in frequency along with the star formation rate at $z\sim 3$).

In contrast, the number counts of highly ionized absorbers---traced principally by the \civ\ doublet---decrease markedly above $z\sim5.5$, though it is challenging to establish how rapid the decline is since \civ\ absorbers are so rare at $z>6$ and require large-pathlength surveys to uncover. As measured by either the number frequency of absorption $dN/dX$, or by its mass contribution to the closure density $\Omega_\mathrm{CIV}$, the prevalence of highly-ionized carbon declines by roughly a factor of 10 from $z=2$ to $z=6$, dropping precipitously at $z\gtrsim5$ (\citealt{2013ApJ...763...37C,2011ApJ...738..159S,2004ApJ...606...92S,2010MNRAS.401.2715D,2013MNRAS.435.1198D,2019MNRAS.483...19M,2018MNRAS.481.4940C}; D. Miller et al. in prep.). Indeed by $z\sim 6$ the frequency of \mgii\ doublets with $W_r>300$m\AA\ exceeds that of \civ\ doublets at the same $W_r$ threshold by a factor of roughly three, opposite to the situation at $z\sim 3$ where \civ\ is much more common.

This reversal in relative comoving number frequencies of low- and high-ionization systems requires a population of absorbers at $z\gtrsim 6$ with no detectable highly-ionized phase. Such systems are rare at $z<5$, where even truly neutral DLAs are typically accompanied by \civ, \siiv, and even \ovi\ absorption. This gas, which is either heated locally or photoionized by the UV background, is usually thought to arise from a more tenuous and enveloping circumgalactic medium in which the cold and neutral clumps are embedded \citep[e.g.,][]{2016ApJ...830...87S}.

While past studies have remarked on these different evolutionary trends by comparing number counts and {\it populations} of low- and high- ionization ions, there has been little systematic investigation of low/high ionization ratios in {\em individual} $z>6$ absorbers. Cosmological hydrodynamic simulations show that studies of absorption lines of multiple elements and ionization states yield insight on both chemical enrichment and the ultraviolet background radiation at the end of reionization \citep{2015MNRAS.447.2526F,2016MNRAS.459.2299F,2018MNRAS.475.4717D}. Such studies require a large parent sample of high-quality infrared spectra to assemble a collection of low-ionization absorbers, and then measure the \civ\ equivalent width (or upper limit) at its expected location. It is also helpful to have spectra of selected objects at high signal-to-noise ratio (SNR) and/or resolution, to gauge whether the aforementioned paucity of \civ\ absorbers at $z\sim6$ reflects systematics in completeness corrections coupled with a decreased ability to detect weaker absorbers in typical IR spectra of distant and faint QSOs, compared to those at lower redshift.

In this work, we present a new sample of absorption line systems at $5.0<z<6.8$, detected in the infrared spectra of 47 quasars at $5.7 < z < 7.5$. A single sightline is also presented with echelle resolution in the red optical and unusually high SNR in the infrared, to study a subset of the parent sample in more detail. Comparing these to various reference samples from the literature and archival spectra at $z=2-4$, we investigate whether:
\begin{itemize}
\item{Low-redshift analogs exist for the high-redshift population of low-ionization absorption systems.}
\item{Unseen carbon could be hidden in an unobservable circumgalactic phase that is non-neutral but has low metallicity.}
\item{The decline in mass density of \civ\ is caused by a change in the ionization state approaching hydrogen reionization, or by a decline in the carbon abundance.}
\end{itemize}
Analysis of high-redshift heavy element absorbers is made more challenging by the complete saturation of the Lyman-$\alpha$ forest, which makes measurement of individual \nhi\ values impossible. This work therefore also explores statistical methods to infer aggregate trends in the physical composition (i.e. metallicity, ionization) of $z>6$ absorbers via extrapolation of \hi\ statistics measured at lower redshift. We adopt throughout a flat $\Lambda$CDM cosmology with ($\Omega_\mathrm{M}$,$\Omega_\Lambda$,$H_0$)=(0.3,0.7,70 \kms\ Mpc$^{-1}$).

\section{Data}

Our absorber sample is drawn from the spectra of 47 $z>5.7$ quasars obtained with the folded-port infrared echellette (FIRE) on the Magellan Baade Telescope \citep{2013PASP..125..270S}. All observations were conducted with a 0.6\arcsec\ slit, providing a spectral resolution of $R=6000$ ($\Delta v=50$ \kms). We have previously used these data to compile completeness-corrected statistics detailing the evolution of \mgii\ \citep{2012ApJ...761..112M,2017ApJ...850..188C} and \civ\ (\citealt{2011ApJ...743...21S}; D. Miller et al. in prep.) absorbers, and study the evolving properties of high-redshift quasars and the neutral fraction of the IGM \citep{2012Natur.492...79S}. All data were recorded in sample-up-the-ramp mode and reduced using the {\tt firehose} pipeline.  This software trims and flat-fields each raw spectral frame, and performs 2D sky subtraction, profile fitting and optimal extraction according to the algorithms of \citet{2003PASP..115..688K}. Individual orders are flux-calibrated and corrected for telluric absorption using contemporaneous observations of A0V stars, as described in \citet{2004PASP..116..362C}. 

The FIRE sample provides a significant survey pathlength to search for rare absorption systems at $z\gtrsim6$. However, many of the spectra have SNR of only 5--10 (i.e. just above threshold for absorption searches), and exhibit spurious positive and negative excursions in flux from residuals of telluric correction or sky subtraction near bright OH lines. At $\Delta v\sim 50$ \kms the metal-line features of interest are generally unresolved.

To capture more detail on a selected subset of absorbers, we therefore supplement the main sample with a Keck/HIRES \citep{1994SPIE.2198..362V} spectrum of the ultraluminous $z=6.3$ quasar J010013.02+280225.8 \citep[hereafter J0100,][]{2015Natur.518..512W}. At $J=17.0$ this object is among the very few high-$z$ quasars amenable to true high-resolution optical spectroscopy. Our HIRES observations were taken with a 0.86\arcsec ~slit yielding $R=50,000$ ($\Delta v=6.7$ \kms), using two different grating angle setups to achieve full wavelength coverage. The total HIRES integration times were 3.8 and 3.0 hours in each setting. This observation was paired with a 7.2-hour FIRE integration to obtain sensitive limits on \mgii, \civ, and other low- and high-ionization species.

We reduced the HIRES data using the \texttt{makee} pipeline, which performs flat fielding, sky subtraction, and order-by-order 1D extraction. We normalized each order with a cubic-spline continuum fit, using manually selected knots. In this region the systematic effect of continuum errors is small compared to Poisson noise in the extracted spectrum. Although there is transmitted flux in the near zone of this quasar, we do not consider absorption lines blueward of \lya\ at the systemic redshift of the QSO, because of greater continuum uncertainty and possible confusion with \hi\ absorption. The normalized single-order spectra from all setups were then coadded using \texttt{makee}. We corrected the resulting 1D spectrum for telluric absorption using a model constructed from contemporaneous observations of a hot white dwarf spectrophotometric standard star.

\section{Identification and measurement of $z>5$ absorption systems}\label{sec:FIRE}

\subsection{Line Identification}

Most, but not all, sightlines studied here were also included in the systematic \mgii\ and \civ\ surveys described above (\citealt{2017ApJ...850..188C}; D. Miller et al. in prep.). We constructed a master list of heavy-element lines, listed in Table \ref{tab:FIREtable}, by first confirming the doublets reported in those papers, and then identifying all other heavy element lines at the reported redshifts of \mgii\ and/or \civ. We next searched the spectra for residual absorption lines not associated with systems reported in these surveys, and manually identified redshifts based on other multi-line coincidences. For sightlines not included in the earlier surveys, we searched first by hand for \mgii\ and \civ\ doublets and then followed an identical procedure using other species. We iterated this procedure until all high-significance lines in each spectrum were classified.

In total there are 63 absorption systems at $z>5$ included in our FIRE sample, 53 of which are identified via \mgii\ or \civ\ doublets customarily used for such surveys. The 10 other absorbers all have \feii\ multiplet absorption; \mgii\ falls in the telluric-line dense H/K bandgap for seven of these, and is impacted by other telluric lines in the remaining 3. While our analysis below centers on absorbers at $z>5$, lower redshift absorption systems were also noted to avoid misidentification.

\subsection{Absorption Measurements}

The column densities reported in Table \ref{tab:FIREtable} are measured using the Apparent Optical Depth method \citep[AOD,][]{1991ApJ...379..245S}, with spectra normalized using a low-order polynomial model of the local continuum ($\Delta\lambda\sim100$\AA\ from the absorption centroid). Non-detections are reported as upper-limits, derived by measuring the 3$\sigma$ upper limit on $W_r$ over one resolution element ($50$ \kms) and converting to column density assuming they are on the linear portion of the curve of growth.

Uncertainties for AOD measurements of unresolved spectral features are not straightforward to assess, since convolution with the spectral response function can lead to underestimated measurements, especially when lines near saturation. The \textit{statistical} uncertainties of the measurements in Table \ref{tab:FIREtable} are typically around 0.05--0.1 dex, although this varies with spectrum quality. To estimate the degree of uncertainty this introduces in our column densities, we compare AOD column densities measured from FIRE data with Voigt profile fits (described below) of the same lines in the J0100 HIRES spectrum. The FIRE AOD column densities are typically 0 to 0.3 dex lower than the HIRES measurements, and we assume this range to be more typical of the inaccuracy present in the column densities measured from the FIRE spectra. While such uncertainties could complicate rigorous analysis of individual systems such as photoionization modeling, the more global statistical analysis presented here is fairly insensitive to these errors.

A more sophisticated measurement procedure is warranted for J0100, where we have fully-resolved optical spectra of several transitions, and high SNR at both optical and IR wavelengths. For this object, we performed model-fitting using a custom-developed Markov-Chain Monte Carlo code that generates Voigt profiles. Written in {\it python} using the {\tt emcee} package \citep{2013PASP..125..306F}, this software jointly solves for the column density and Doppler parameter of all ionic transitions at each fitted redshift, naturally producing upper limits for non-detections. It also provides a check against saturation, which manifests as a high-end tail of the posterior column density distribution.

The model assumes that all low-ionization lines share a common temperature and characteristic random turbulent velocity distribution, and fits for these parameters along with column density. As expected, the model fits performed well for the largely unsaturated metal-line profiles in HIRES. We measure ions with transitions only at wavelengths beyond HIRES' spectral range (e.g., \mgii, \feii) with fits to the FIRE data. FIRE's much broader line-spread function prevents detailed study of velocity structure and creates clearly visible parameter degeneracies in the MCMC fit posteriors when multiple redshift components are used. To improve the FIRE modeling, we performed two tests. First, we fit the FIRE data with a single-component Voigt profile using no prior information from HIRES. Then, we fit a model with multiple components whose redshifts and Doppler parameters were fixed to the values output by the HIRES fit of other low-ionization transitions (e.g. \cii), whose velocity structure we regarded as ground truth. We further constrained the ratio of column densities for each redshift component of a given ion to be the same for all ions, limited to a narrow range bounded by the ions fit in HIRES.

These two methods yielded consistent total column densities at the $0.1-0.2$ dex level, and results from the latter approach are reported in Table \ref{tab:J0100abs}. The fitted values are in tight agreement with AOD measurements made with HIRES, but systematically higher than AOD measurements on the unresolved FIRE data. This is expected for unresolved and/or mildly saturated lines, since convolution with the line-spread function kernel distributes power to the profile wings where the non-linear nature of the AOD conversion leads to slight underestimates. In subsequent analysis we use the MCMC-fitted values rather than AOD measurements for J0100.

\subsection{Comparison with Deep Spectra from the Literature}

One of our sightlines---ULAS J1120+0641 at $z=7.08$---was observed separately by \citet{2017MNRAS.470.1919B} for 30 hours using VLT/X-Shooter. These data have similar spectral coverage and resolution as FIRE, but higher SNR and better telluric correction on account of their long exposure time. This provides an opportunity for informed comparisons on the role of data quality in our derived measurements and scientific conclusions.

We detect two of the seven intervening absorbers they identify at $z>5$. At $z=5.795$ we both detect only \civ, with comparable column densities. At $z=5.508$, we measure similar column densities for \siii\ and \feii, but a higher value of \N{CIV} inconsistent with their upper-limit. Only the 1548\AA\ transition of the \civ\ doublet is detected at $3\sigma$ in our spectrum, so this is quite possibly noise contamination. Of the five remaining absorbers, four are contaminated by telluric absorption in our spectrum and one (at $z=6.407$) is below the detection threshold of FIRE data (in \cii\ and \mgii) with $\rm{SNR}\sim10$. We include these five absorbers in figures, but not in statistical calculations.

We note there are several additional instances where higher SNR spectra from the literature show weak detections of ions we do not detect. For example, we only detect low-ionization species in the $z=5.065$ absorber in sightline SDSS J0818+1722, whereas \citet{2019MNRAS.483...19M} detect \civ\ in a spectrum with SNR$>100$. Similarly, \citet{2018MNRAS.481.4940C} detect \cii\ at $z=5.574$ in ULAS 1319+0950, where we only detect \cii. In all instances our upper limits are 0.2-0.3 dex below column densities measured from other spectra. We opt to use the upper limits we obtained for these systems, to maintain uniformity in the FIRE sample. Using the detected column densities from the literature would slightly change some of the statistics discussed below, but comparisons between species are fairly robust, since the ions we do detect in these instances have markedly higher column densities.

\startlongtable
\begin{deluxetable*}{lllllllllll}
\tabletypesize{\footnotesize}
\tablecaption{$z>5$ FIRE Absorbers\label{tab:FIREtable}}
\tablehead{\colhead{Quasar Name\tablenotemark{a}} & \colhead{$z_\mathrm{abs}$} & \colhead{$\log\N{SiII}$} & \colhead{$\log\N{CII}$} & \colhead{$\log\N{MgII}$} & \colhead{$\log\N{FeII}$} & \colhead{$\log\N{OI}$} & \colhead{$\log\N{CIV}$} & \colhead{$\log\N{SiIV}$} & \colhead{$\log\N{AlII}$} & \colhead{$\log\N{AlIII}$}}
\startdata
ULAS J1342+0928$^{(1)}$ & 6.843 & 13.9    & $>$14.6 & $>$13.8 & 13.6    & ---     & $<$13.5 & $<$12.9 & 12.1    & ---      \\
PSO J231-20$^{(2)}$     & 6.476 & $<$13.2 & 13.5    & $<$12.3 & $<$12.7 & $<$14.0 & 14.0    & 13.3    & ---     & ---      \\
VIK J2316-2802$^{(3)}$  & 6.470 & $<$13.4 & $<$13.7 & $<$12.8 & $<$12.6 & $<$14.3 & 13.6    & $<$13.0 & $<$12.5 & ---      \\
PSO J323+12$^{(2)}$     & 6.447 & $<$12.3 & $<$13.3 & 12.4    & $<$12.3 & $<$13.8 & $<$13.1 & $<$12.7 & ---     & ---      \\
ULAS J1342+0928         & 6.271 & 13.8    & ---     & 13.5    & 13.3    & ---     & $<$13.8 & ---     & 11.6    & ---      \\
VDES J0224-4711$^{(4)}$ & 6.269 & 12.7    & $<$13.7 & 12.8    & $<$12.3 & ---     & $<$13.5 & $<$12.9 & 12.4    & ---      \\
VIK J2348-3054$^{(5)}$  & 6.268 & $<$13.9 & 14.0    & 13.4    & 13.3    & ---     & $<$13.7 & $<$13.8 & 12.6    & ---      \\
PSO J159-02$^{(6)}$     & 6.238 & 13.2    & 14.1    & 13.1    & 13.3    & 14.5    & 14.1    & 13.5    & $<$12.2 & $<$12.8  \\
VIK J1048-0109$^{(7)}$  & 6.221 & $>$14.6 & $>$14.8 & $>$14.2 & 14.1    & $>$15.1 & $<$13.4 & $<$13.5 & 13.0    & $<$13.0  \\
VHS J0411-0907$^{(8)}$  & 6.178 & 13.7    & 14.1    & ---     & 13.3    & ---     & $<$13.3 & $<$13.2 & 12.4    & $<$12.6  \\
VDES J0224-4711         & 6.123 & $<$13.2 & $<$13.6 & 12.9    & 12.8    & ---     & $<$13.5 & $<$12.8 & 12.1    & $<$12.7  \\
PSO J183+05$^{(2)}$     & 6.064 & 13.9    & 14.4    & 13.5    & 12.9    & 14.4    & 14.4    & 13.8    & $<$12.1 & $<$12.5  \\
PSO J159-02             & 6.055 & 13.1    & 14.2    & 13.3    & 12.7    & 14.2    & $<$13.2 & $<$13.1 & $<$12.5 & $<$12.8  \\
SDSS J2310+1855$^{(9)}$ & 5.938 & 13.0    & 13.6    & 12.8    & 13.0    & 14.2    & $<$12.8 & $<$12.5 & ---     & $<$12.2  \\
VHSJ0411-0907           & 5.936 & 13.8    & ---     & ---     & 13.4    & ---     & $<$13.4 & $<$12.7 & $<$12.9 & $<$12.8  \\
CFQS1509-1749$^{(10)}$  & 5.916 & $<$12.1 & $<$13.3 & ---     & $<$12.6 & ---     & 14.0    & 13.1    & $<$11.9 & $<$12.4  \\
PSO J159-02             & 5.913 & 13.8    & 14.5    & ---     & 13.0    & ---     & $<$13.7 & $<$13.0 & 12.6    & ---      \\
ULAS J1342+0928         & 5.889 & ---     & ---     & ---     & $<$13.2 & ---     & 14.1    & ---     & $<$12.7 & $<$13.5  \\
PSO J183+05             & 5.844 & 13.4    & 14.4    & ---     & 13.3    & ---     & 13.4    & $<$13.2 & $<$12.1 & $<$13.3  \\
ULAS J1120+0641$^{(11)}$& 5.795 & ---     & ---     & ---     & $<$12.5 & ---     & 14.0    & ---     & ---     & $<$12.5  \\
SDSS J1411+1217$^{(12)}$& 5.787 & $<$13.2 & $<$13.5 & ---     & $<$12.3 & $<$14.4 & 14.3    & ---     & $<$12.3 & $<$12.8  \\
PSO J247+24$^{(2)}$     & 5.785 & 14.4    & ---     & ---     & 13.4    & ---     & 13.7    & ---     & $<$13.0 & $<$13.1  \\
SDSS J1030+0524$^{(13)}$& 5.744 & 14.1    & 14.7    & ---     & 13.5    & ---     & 14.1    & 14.1    & ---     & $<$12.9  \\
PSO J159-02             & 5.734 & 14.1    & 14.4    & ---     & 13.5    & ---     & $<$13.2 & ---     & ---     & $<$12.5  \\
SDSS J1030+0524         & 5.725 & $<$13.3 & ---     & ---     & $<$12.4 & ---     & $>$14.5 & ---     & ---     & $<$12.4  \\
DES0454-4448$^{(14)}$   & 5.697 & $<$13.4 & ---     & ---     & $<$12.5 & $<$13.7 & 13.9    & 13.5    & 12.5    & $<$12.7  \\
PSO J209-26$^{(6)}$     & 5.635 & 14.1    & $>$14.7 & ---     & 13.6    & 14.8    & 14.0    & 13.3    & 12.9    & $<$12.4  \\
PSO J217-07$^{(6)}$     & 5.630 & 13.9    & $<$14.2 & ---     & 13.9    & ---     & $<$13.9 & $<$13.3 & 12.7    & $<$12.6  \\
ULAS 1319+0950$^{(15)}$ & 5.574 & 12.9    & $<$13.7 & ---     & 12.8    & ---     & 14.0    & 13.6    & 12.3    & $<$12.9  \\
PSO J323+12             & 5.519 & ---     & ---     & ---     & $<$12.3 & ---     & 13.8    & ---     & $<$12.0 & $<$12.5  \\
SDSS J1030+0524         & 5.517 & 13.7    & ---     & ---     & 13.2    & ---     & 14.0    & 13.4    & ---     & $<$12.5  \\
PSO J247+24$^{(2)}$     & 5.511 & $<$13.7 & ---     & ---     & 13.6    & ---     & $<$13.6 & ---     & $<$12.7 & $<$12.9  \\
ULAS J1120+0641         & 5.508 & 13.3    & ---     & ---     & 13.0    & ---     & 13.3    & ---     & $<$12.0 & $<$12.5  \\
SDSS J2310+1855         & 5.487 & $<$13.0 & $<$13.1 & ---     & $<$12.0 & ---     & 13.8    & 12.9    & ---     & $<$12.4  \\
PSO J213-22$^{(16)}$    & 5.462 & 13.5    & $<$13.5 & ---     & $<$13.1 & ---     & 14.1    & $<$13.2 & 12.3    & $<$12.8  \\
PSO J209-26             & 5.415 & $<$13.8 & $<$13.8 & ---     & $<$13.0 & ---     & 14.0    & 13.2    & $<$12.2 & $<$12.4  \\
PSO J323+12             & 5.362 & $<$13.4 & ---     & $<$12.0 & $<$12.6 & ---     & 14.0    & ---     & $<$12.1 & $<$12.2  \\
PSO J217-16$^{(6)}$     & 5.357 & 14.5    & ---     & $>$14.2 & 14.0    & ---     & 14.5    & 13.8    & 13.5    & 13.2     \\
PSO J036+03$^{(17)}$    & 5.354 & $<$13.4 & ---     & 12.8    & 13.0    & ---     & $<$13.2 & ---     & ---     & $<$12.4  \\
SDSS J1411+1217         & 5.331 & $<$13.4 & ---     & 12.9    & 12.6    & ---     & $<$13.8 & $<$13.3 & $<$12.0 & $<$12.5  \\
PSO J239-07$^{(6)}$     & 5.324 & $<$13.4 & ---     & 12.9    & $<$13.0 & ---     & ---     & $<$12.9 & 12.6    & $<$12.4  \\
SDSS J0836+0054$^{(13)}$& 5.323 & $<$12.9 & $<$13.0 & $<$11.7 & $<$12.7 & ---     & 13.3    & 13.0    & $<$11.6 & $<$12.0  \\
ATLAS J025-33$^{(18)}$  & 5.316 & ---     & ---     & 13.6    & 13.6    & ---     & 14.2    & ---     & 12.7    & $<$12.7  \\
SDSS J2310+1855         & 5.288 & $<$12.8 & ---     & $<$11.5 & $<$12.2 & ---     & 13.6    & $<$13.0 & $<$11.6 & $<$12.1  \\
PSO J209-26             & 5.276 & 13.9    & 13.8    & 13.0    & 12.9    & ---     & $<$13.3 & $<$12.9 & 12.2    & $<$12.7  \\
PSO J183-12$^{(16)}$    & 5.272 & $<$13.0 & ---     & $<$11.8 & $<$12.3 & ---     & 13.5    & $<$13.1 & 12.5    & $<$12.6  \\
SDSS J1411+1217         & 5.250 & $<$13.5 & ---     & 12.9    & $<$12.5 & ---     & 14.1    & 13.5    & $<$12.1 & $<$13.0  \\
PSO J239-07             & 5.245 & ---     & ---     & 12.5    & $<$12.7 & ---     & 13.7    & ---     & $<$12.1 & $<$12.6  \\
PSO J036+03             & 5.243 & $<$13.4 & ---     & $<$12.8 & $<$12.6 & ---     & 14.0    & ---     & $<$12.0 & $<$12.6  \\
PSO J217-07             & 5.225 & ---     & ---     & 12.9    & 13.4    & ---     & $<$13.5 & ---     & $<$12.8 & ---      \\
VIK J2348-3054          & 5.221 & ---     & ---     & $<$12.5 & $<$12.8 & ---     & 14.4    & ---     & $<$12.6 & ---      \\
PSO J209-26             & 5.202 & 14.3    & ---     & 13.3    & 13.8    & ---     & 13.7    & $<$13.0 & 12.8    & $<$12.6  \\
ATLAS J025-33           & 5.189 & $<$13.3 & ---     & $<$11.8 & $<$12.2 & ---     & 13.3    & ---     & $<$11.8 & $<$12.3  \\
PSO J071-02$^{(6)}$     & 5.174 & 14.1    & ---     & 13.9    & 13.3    & ---     & 14.3    & 13.8    & 12.9    & 13.2     \\
SDSS J0836+0054         & 5.126 & $<$13.1 & ---     & $<$11.9 & $<$12.0 & ---     & 13.7    & 13.2    & $<$12.0 & $<$12.5  \\
PSO J239-07             & 5.120 & ---     & ---     & 12.9    & 12.5    & ---     & $<$13.7 & ---     & $<$12.4 & $<$13.0  \\
VDESJ0224-4711          & 5.109 & $<$13.5 & ---     & ---     & $<$12.6 & ---     & $>$14.8 & ---     & $<$12.4 & ---      \\
SDSS J0818+1722$^{(19)}$& 5.065 & 13.6    & ---     & $>$13.4 & 13.3    & ---     & $<$13.1 & $<$12.7 & 12.2    & ---      \\
SDSS J0842+1218$^{(20)}$& 5.048 & 14.4    & ---     & $>$13.8 & 14.1    & ---     & ---     & ---     & 12.8    & $<$13.1  \\
PSO J247+24$^{(2)}$     & 5.039 & 13.9    & ---     & 13.5    & ---     & ---     & ---     & ---     & ---     & ---      \\
PSO J159-02             & 5.022 & 13.6    & ---     & $<$12.6 & $<$13.0 & ---     & 14.3    & ---     & $<$12.3 & 13.5     \\
VDESJ0224-4711          & 5.006 & $<$13.0 & ---     & 12.9    & $<$12.5 & ---     & 14.2    & ---     & 12.7    & ---      \\
VIK J0109-3047$^{(4)}$  & 5.001 & ---     & ---     & 13.0    & $<$12.8 & ---     & ---     & ---     & ---     & --- 
\enddata
\tablenotetext{a}{Superscript numbers correspond to quasar discovery references.}
\tablecomments{Column densities are in units of cm$^{-2}$. Upper limits are 3$\sigma$, measured across one spectral resolution element.}

\tablerefs{(1) \citet{2018Natur.553..473B}, (2) \citet{2017ApJ...849...91M}, (3) B. Venemans, private communication , (4) \citet{2017MNRAS.468.4702R}, (5) \citet{2013ApJ...779...24V}, (6) \citep{2016ApJS..227...11B}, (7) \citet{2017ApJ...839...27W}, (8) \citet{2018arXiv181011926W}, (9) \citet{2016ApJS..227...11B}, (10) \citet{2007AJ....134.2435W}, (11) \citet{2011Natur.474..616M}, (12) \citet{2004AJ....128..515F}, (13) \citet{2001AJ....122.2833F}, (14) \citet{2015MNRAS.454.3952R}, (15) \citet{2009AA...505...97M}, (16) \citet{2014AJ....148...14B}, (17) \citet{2015ApJ...801L..11V}, (18) \citet{2015MNRAS.451L..16C}, (19) \citet{2006AJ....131.1203F}, (20) \citet{2015AJ....149..188J}}
\end{deluxetable*}

\begin{figure}[h!]
\includegraphics[width=\linewidth]{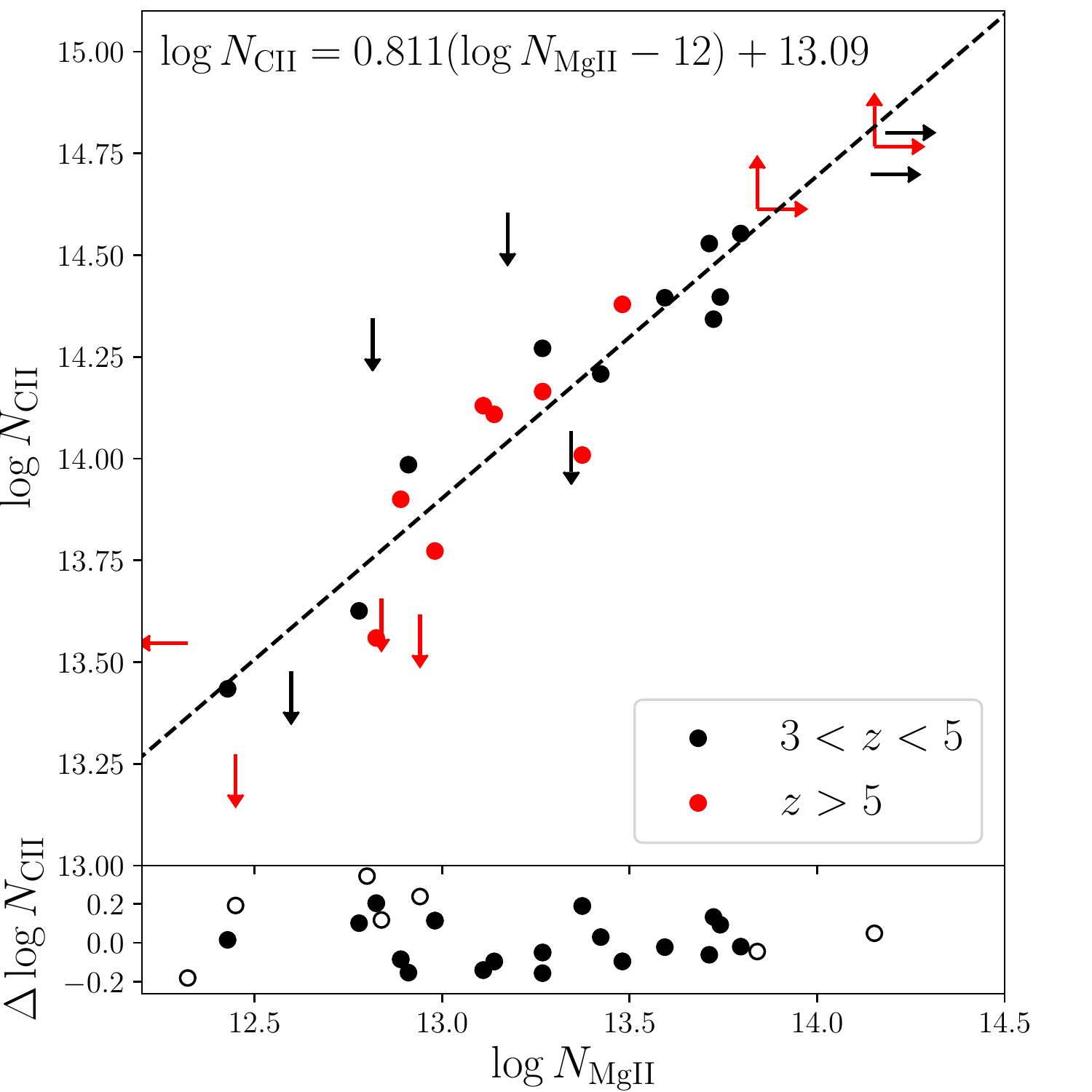}
\caption{
  Column densities of singly-ionized carbon and magnesium. Red points are our measurements at high-redshift, and black points are measured from $z=3-5$  \mgii\ absorbers identified in \citet{2017ApJ...850..188C}. Arrows and open circles in the residuals indicate when at least one of the two species is undetected or saturated. There is a clear correlation between \N{CII} and \N{MgII}, and no obvious difference with redshift in the range considered. The dotted line is a linear fit to all points where both species are detected, with residuals shown below (unfilled circles correspond to limits). All detections are within $\pm0.2$ dex of the fit. The correlation between \cii\ and \mgii\ enables us to convert \mgii\ measurements to \cii\ estimates. \label{fig:c2mg2}}
\end{figure}

\begin{figure*}[t!]
\includegraphics[width=\linewidth]{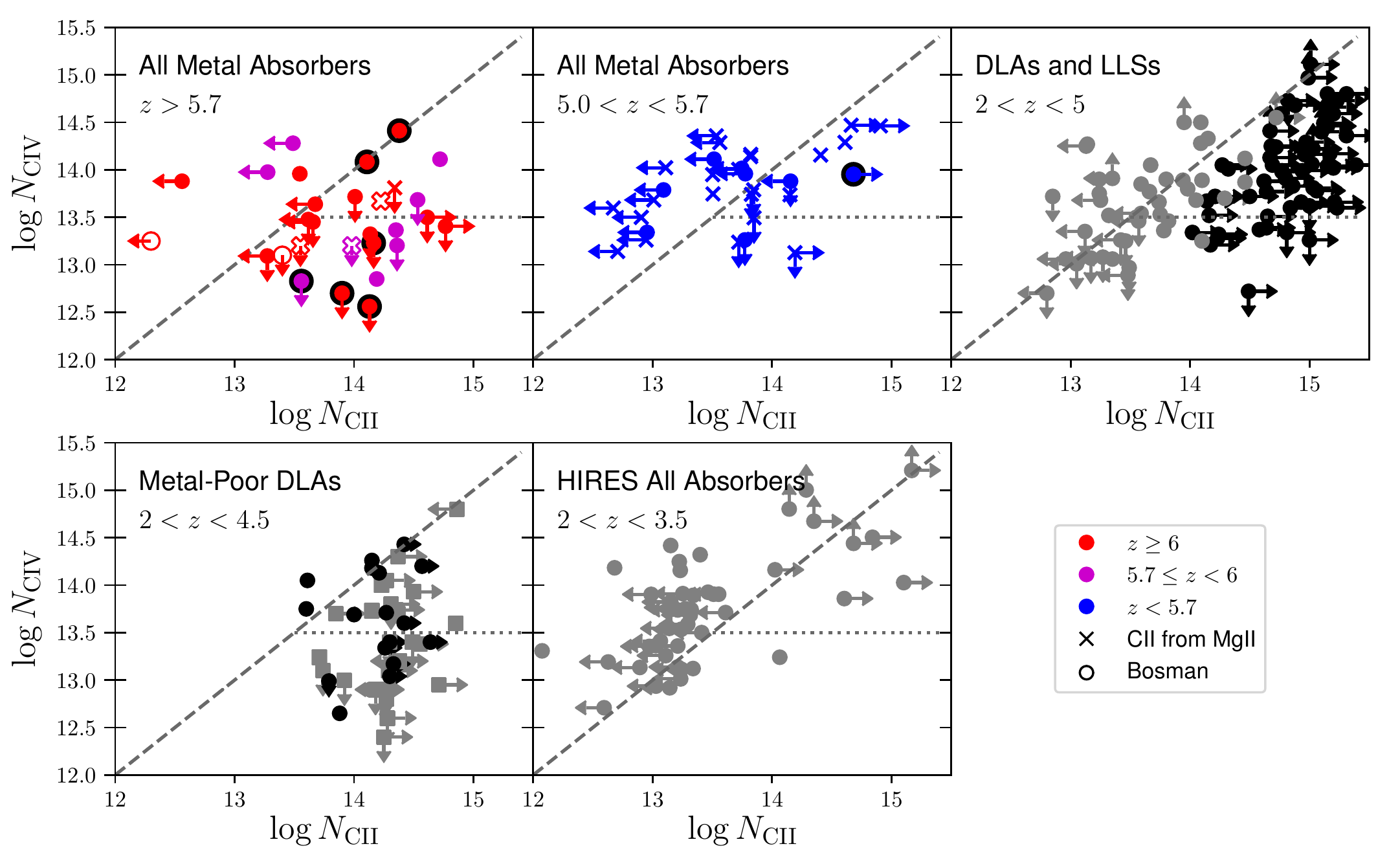}
\caption{Top left \& middle: Column densities of singly- and triply-ionized carbon in high redshift absorption systems, and comparison samples. Crosses indicate absorption systems where \N{CII} is estimated from \N{MgII}. Absorbers are binned by color into low ($5<z<5.7$), intermediate ($5.7<6$), and high $z>6$ redshift. Unfilled points are from \citet{2017MNRAS.470.1919B}. Black circles surround absorbers where \oi\ is detected; none of the absorbers where \cii\ is detected have \oi\ nondetections, but rather \oi\ is often inaccessible. The dashed gray line indicates equality between the two ions, and the dotted line at at $\log\N{CIV}=13.5$ isolates the portion of parameter space sparsely filled at lower redshift. Top right: DLAs at $2<z<5$ (black) and LLSs at $z\sim3.5$ (gray). Bottom: DLAs with [O/H]$<-2$ (black circles are from Table \ref{tab:MPDLAs}, gray squares are from \citealp{2010ApJ...721....1P}) and all absorbers identified in a search of $z\sim3$ HIRES spectra (see Section \ref{sec:analogs}). Comparing to lower redshift absorbers, $z>5.7$ metal absorbers appear most similar to metal-poor DLAs.} 
\label{fig:highz_ratios}
\end{figure*}

\subsection{Correlations in Low-Ionization Absorption, and their Utility as Proxies}

Ideally, ionization analyses utilize level ratios of the same element to avoid scatter resulting from variations in relative abundances or differential depletion. In much of our analysis we consider the ratio $\N{CII}/\N{CIV}$. However in practice this is not always possible because 
the low-ionization \cii\ $\lambda1334$\AA\ line falls within the saturated \lya\ forest at large $\Delta z$ from the QSO emission redshift.

In these cases it is often possible to measure \mgii, and in select examples both ions can be measured in the same system because of FIRE's broad wavelength coverage. Figure \ref{fig:c2mg2} presents measurements of these systems compiled from Tables \ref{tab:FIREtable} and \ref{tab:J0100abs} (red points), combined with similar FIRE measurements at $3<z<5$ (black points) from absorbers identified in \citet{2017ApJ...850..188C}. A strong correlation is evident that we fit (to both sets of points concurrently) with a simple linear regression, including only points where both ions are measured:

\begin{equation}
\log(\N{CII})=[0.81\pm0.07](\log(\N{MgII})-12)+[13.09\pm0.10] \label{eq:c2mg2}
\end{equation}
with $\pm0.2$ dex residual scatter. Over the $\log(\N{MgII})=12.5-14.0$ range encompassing our sample, the corresponding ratio $\log(\N{CII}/\N{MgII})=1.0-0.7$ may be compared with the Solar relative [C/Mg] abundance of 0.83 dex \citep{2009ARA&A..47..481A}, indicating that such systems could plausibly have only modest variation in ionization and roughly Solar relative abundances. In this picture the slight deviation from unity slope could reflect either an increasing fractional dust depletion of carbon amongst stronger \mgii\ absorbers, or a slow change in their ionization.

In the discussion below, we use \mgii\ as a proxy for \cii\ in systems
where the $\log(\N{CII}/\N{CIV})$ ratio cannot be measured directly, using
Equation \ref{eq:c2mg2} to convert between ions. For all such cases,
the \mgii-derived measurements or limits are marked separately in
figures to distinguish from single-species measurements. When we
detect \mgii\ but find an upper-limit for \cii, we retain the
upper-limit in our analysis.

\section{Results}

\subsection{Ratios of Low- to High-Ionization Carbon\label{sec:ratios}}

Figure \ref{fig:highz_ratios} shows measurements of \cii\ (or the implied \cii\ from \mgii) and \civ\ for 54 individual $z>5$ absorbers from Tables \ref{tab:FIREtable} and \ref{tab:J0100abs} and five from \citet{2017MNRAS.470.1919B}. The top-right panel displays similar ratios for DLAs \citep{1999ApJS..121..369P,2001ApJS..137...21P,2003ApJS..147..227P,2007ApJS..171...29P} and LLSs \citep{2015ApJ...812...58C,2016ApJ...833..270G} at lower redshifts, between $2.7<z<4.7$.

The first qualitative result is that the majority of systems at $z>5.0$ contain upper limits in either \cii\ or \civ---34 of 55 absorbers ($62\%$) fit this description, with 17 systems (31\%) detected in each carbon state and not the other. Seventeen systems exhibit both high- and low-ionization carbon, and four are undetected in both species (having been identified by \mgii\ or other transitions).  

A separate examination of the points by redshift reveals a trend in the sense of the limits, such that the highest redshift part of the sample is more likely to be detected only in \cii; there is a paucity of \civ\ absorbers at the highest redshifts. A dashed line in the figure denotes the $\N{CII}=\N{CIV}$ locus, and the dotted line at $\log\N{CIV}=13.5$ isolates the region that is sparsely filled at lower redshift.

Dividing the sample at $z=5.7$ based on an apparent qualitative change in absorber properties at this redshift (and excluding absorbers with upper-limits to both ions), we find that 11 of 21 ($52\pm10\%$) of the highest-redshift systems exhibit only low-ionization absorption; of the remaining systems, five have $\N{CII}\gtrsim\N{CIV}$, and five have stronger (or only) \civ. The additional absorbers from \citep{2017MNRAS.470.1919B} have similar characteristics, with three out of five having only low-ionization lines, one having only \civ, and the fifth having both species present but stronger absorption from low ions. A number of other low-ionization systems have been detected at $z\gtrsim6$ \citep{2011ApJ...735...93B}, but do not have high-quality IR spectra covering \civ\ transitions.

At $z<5.7$ the situation is reversed: of 30 total systems, 13 are detected only in \civ, and only six systems ($20\pm7\%$) are \cii-only. This change is consistent with the well-known decrease in the \civ\ mass density $\Omega_\mathrm{CIV}$ \citep{2011ApJ...738..159S,2010MNRAS.401.2715D,2017MNRAS.470.1919B}.

The parameter space occupied by $z>5.7$ absorbers---with strong but unsaturated (or mildly saturated) \cii\ yet weak or undetected \civ---is sparsely populated by heavy element absorbers at $z<5$.  A comparison with \hi-selected DLAs at $2<z<5$ (right panel, see \S \ref{sec:analogs}) illustrates that \cii\ is almost universally saturated given typical DLA metallicities at lower redshifts, and the vast majority also have robust \civ\ with $\log\N{CIV}\gtrsim 13.50-13.75$. LLS absorbers at $z\sim3.5$ have \cii\ column densities similar to our $z>6$ systems, but for $\log\N{CII}\gtrsim 13.50-14.50$, one almost always detects \civ, and LLS ionization models can reproduce concurrent observations of both species when the ionization parameter\footnote{The ionization parameter, $U=n_\gamma/n_\mathrm{H}$, is the ratio of hydrogen-ionizing photon density to hydrogen density.} is in the range $-3<\log U<-2$ \citep[e.g.][]{2016ApJ...833..270G}.

\begin{figure}
\includegraphics[width=\linewidth]{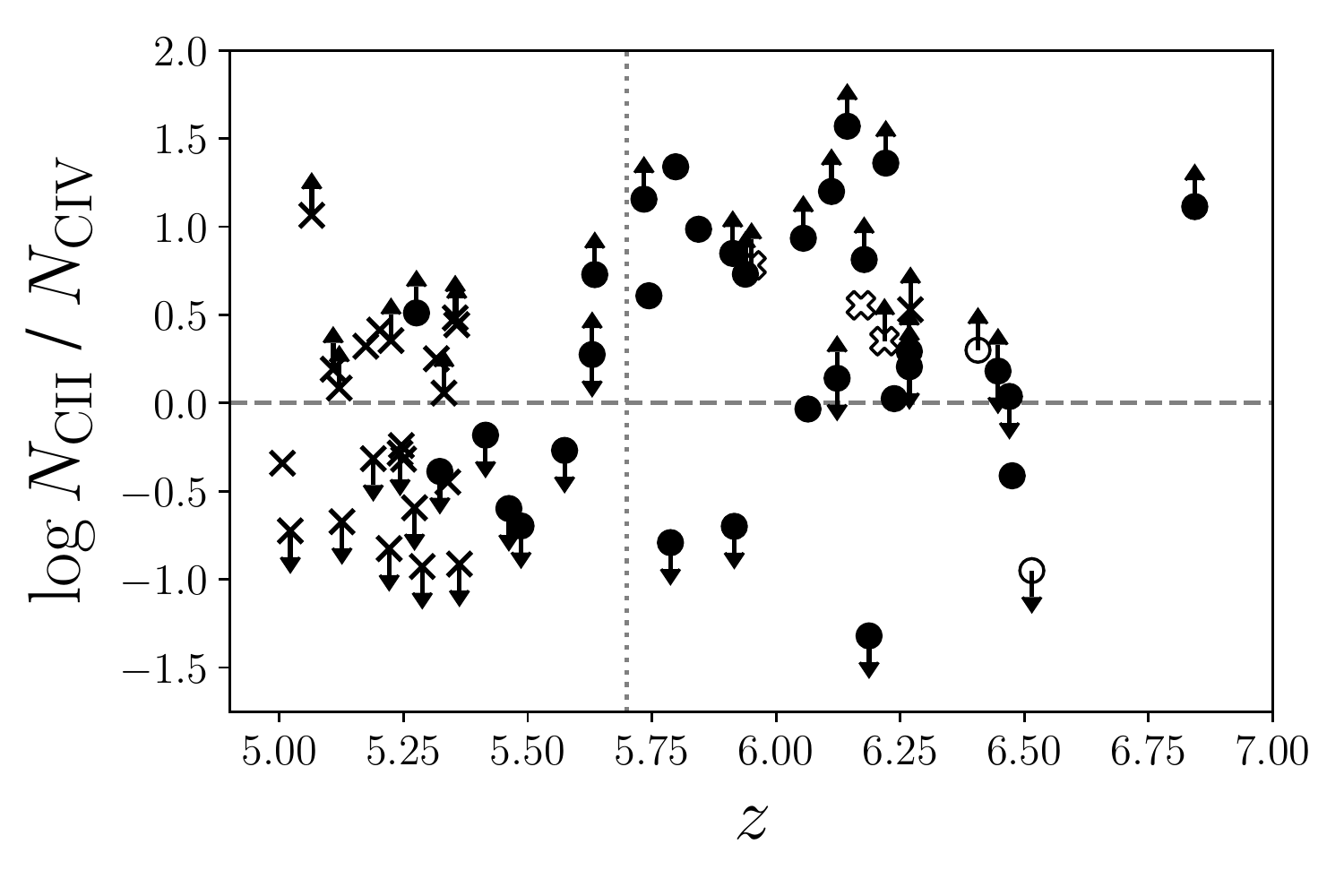}
\caption{Ratio of singly- to triply-ionized carbon. As in
  Figure \ref{fig:highz_ratios}, crosses indicate \N{MgII} converted to \N{CII}, and open markers are from \citet{2017MNRAS.470.1919B}. The horizontal line is at equality and the vertical line is at $z=5.7$. At $z<5.7$ absorbers are roughly evenly split between low- and high-ionization, whereas low-ionization dominates at higher redshifts. There are numerous absorbers at $z>5.7$ that have at least 10 times more \cii\ than \civ.\label{fig:xc_highz}}
\end{figure}

This differentiation is made clearer by plotting the \cii-to-\civ\ ratio as a function of redshift (Figure \ref{fig:xc_highz}). Below $z=5.7$, absorption systems are roughly evenly split between those dominated by low-ionization and high-ionization gas whereas high-ionization absorbers become increasingly rare relative to low-ionization absorbers at $z>5.7$.

At $z\sim 3$, absorbers exhibiting \cii\ without attendant \civ\ are rare but not unprecedented; they tend to be found in DLAs selected specifically for study because of low metallicity. Several groups have developed triage methods to identify such systems with [O/H]$\lesssim -2.5$; these represent the lowest $\sim 10\%$ of the DLA metallicity distribution. 

The lower-left panel of Figure \ref{fig:highz_ratios} illustrates that the metal-poor DLAs (see Section \ref{sec:analogs}) do overlap in parameter space with the low-ionization absorbers seen at $z>5.7$. This raises the possibility that the large majority of heavy-element absorbers at $z>5.7$ are neutral and metal-poor. This conclusion is supported by the detection of \oi\ in every case (unfortunately seldom) where the $1302$\AA\ line of an absorber with other low-ionization species falls in an observable window. Systems with \oi\ detections are outlined with black circles in the upper-left panel of Figure \ref{fig:highz_ratios} for reference.

The suppression of \civ\ at high redshift is striking because \civ\ is nearly ubiquitous in $z\sim 3$ absorption systems, and is thought to arise from highly ionized, warm and tenuous matter in galactic halos, at several tenths of the virial radius \citep{2001ApJ...556..158C,2014ApJ...796..136B}. We will address below in Section \ref{sec:civ} whether the decrease in \civ\ is consistent with lower heavy-element abundances in circumgalactic gas, or a change in ionization conditions \citep{2015MNRAS.447.2526F}, or some combination of these effects. First, we present data on several exemplar systems with fully resolved, high-SNR spectroscopy to inform the reader's intuition about their observed properties.

\subsection{Resolved Absorption Systems at $z\sim6$}\label{sec:J0100}

\begin{deluxetable*}{cccccccc}[t]
\tablecaption{Absorbers along the line of site to QSOJ0100+2802\label{tab:J0100abs}}
\tablehead{
\colhead{$z_{\text{abs}}$} & \colhead{$\log\N{CII}$} & \colhead{$\log\N{SiII}$} &\colhead{$\log\N{OI}$} & \colhead{$\log\N{SiIV}$}  &\colhead{$\log\N{CIV}$} &\colhead{$\log\N{FeII}$} &\colhead{$\log\N{MgII}$\tablenotemark{a}}}
\startdata
6.1873 & $<$12.56 & 11.59\tablenotemark{b} & $<$13.25 & 12.87\tablenotemark{a} & 13.88\tablenotemark{a} & $<$12.25\tablenotemark{a} & $<$11.80\\
6.1435 & 14.13\tablenotemark{c} & 13.39 & 14.72\tablenotemark{c} & $<$12.67\tablenotemark{d} & $<$12.56\tablenotemark{a} & 12.92\tablenotemark{a} & 13.11\\
6.1117 & 13.90 & 12.94 & 14.43 & $<$11.99 & $<$12.70\tablenotemark{a} & 12.55\tablenotemark{a} & 12.89\\
5.7979 & 14.19\tablenotemark{c} & 13.57 & --- & $<$12.12 & 12.85\tablenotemark{a} & 13.10\tablenotemark{a} & ---\\
5.3390 & --- &$<$13.02 & --- &13.18 & 13.95 & 12.46\tablenotemark{a} & 12.51\\
5.1083 & --- & 14.35 & --- & & 14.47 & 13.97 & $>$13.94\\
4.8750 & --- &$<$12.42 & --- & & 13.14 & $<$11.77\tablenotemark{a} & $<$11.53
\enddata
\tablenotetext{a}{Measured from FIRE data}
\tablenotetext{b}{For nondetections over the same range used for \cii\ and \oi\ we find $\log\N{Si II}<12.12$}
\tablenotetext{c}{Mildly saturated}
\tablenotetext{d}{HIRES sensitivity is low here; higher SNR FIRE data yield a limit of $<12.08$}
\end{deluxetable*}

\begin{figure*}[t]
\includegraphics{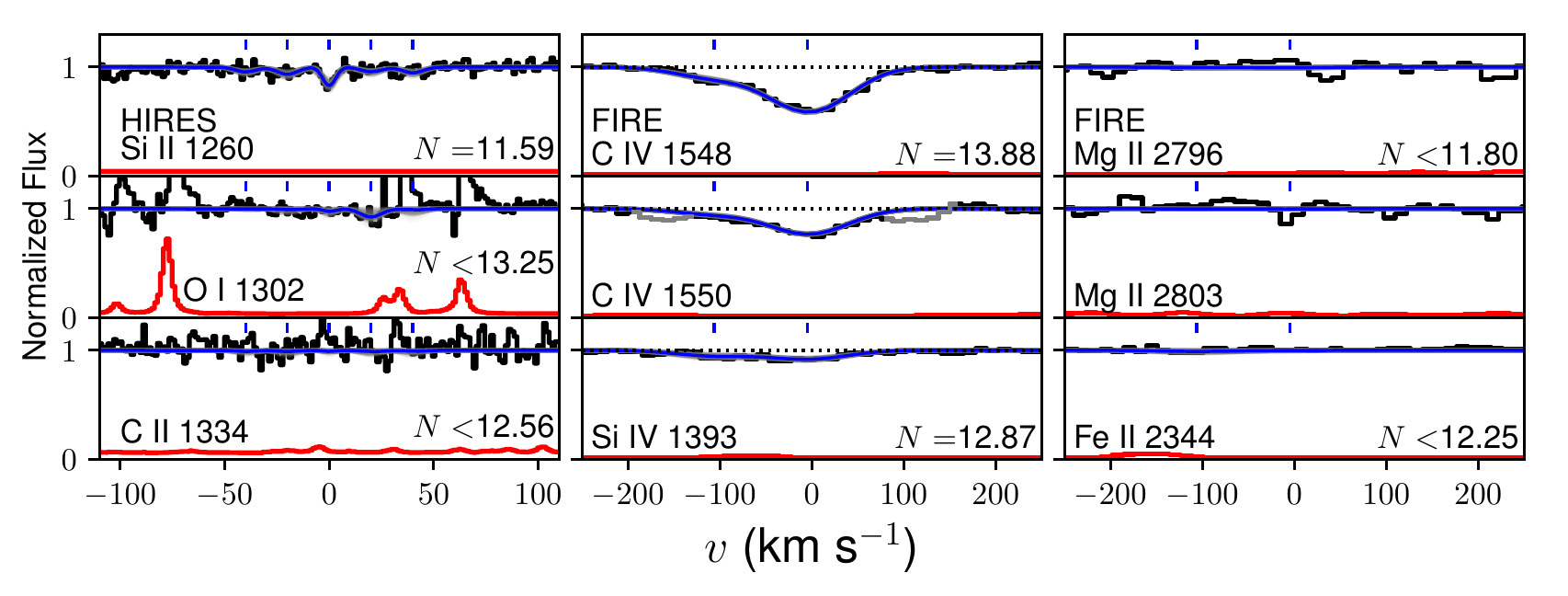}
\caption{Absorption system at $z=6.1873$ absorber along the line of sight to J0100+2802. This absorber, with \civ\ absorption but no \cii\ and weak \siii, is atypical at $z\sim6$ but is akin to commonplace absorbers at $z\lesssim5$. The black and red histograms show the normalized spectrum and $1-\sigma$ errors, respectively; grayed out portions are not fit and correspond to intervening absorption or errors due to, e.g., poor telluric corrections. The blue curve is the Voigt profile fit, overlaid with 100 Voigt profiles with parameters randomly drawn from the posterior overlaid as gray curves. Blue dashes indicate the redshifts of different Voigt profile components. Note the HIRES spectra are shown over a narrower velocity range, as less detail can be seen at the scale needed to display the FIRE spectra. \label{fig:zabs6.18}}
\end{figure*}

The majority of systems presented above are unresolved at FIRE's $\Delta v=50$ km s$^{-1}$, and very few have been observed at full resolution because most quasars at $z>6$ are not bright enough for high-resolution spectroscopy. In the largest high-resolution study to date \citep{2011ApJ...735...93B}, four sightlines were observed with HIRES at moderate SNR, resulting in no detections for three sightlines but {\em four} $z>6$ \oi\ absorbers toward the brightest object, SDSS J1148+5251. All have unsaturated low-ionization absorption (including \cii) without high-ionization lines, similar to our sample, but the corresponding upper limits on \civ\ are less constraining because the required IR spectrum (taken with Keck/NIRSPEC) has much lower SNR than our FIRE sample.

\begin{figure*}
\includegraphics{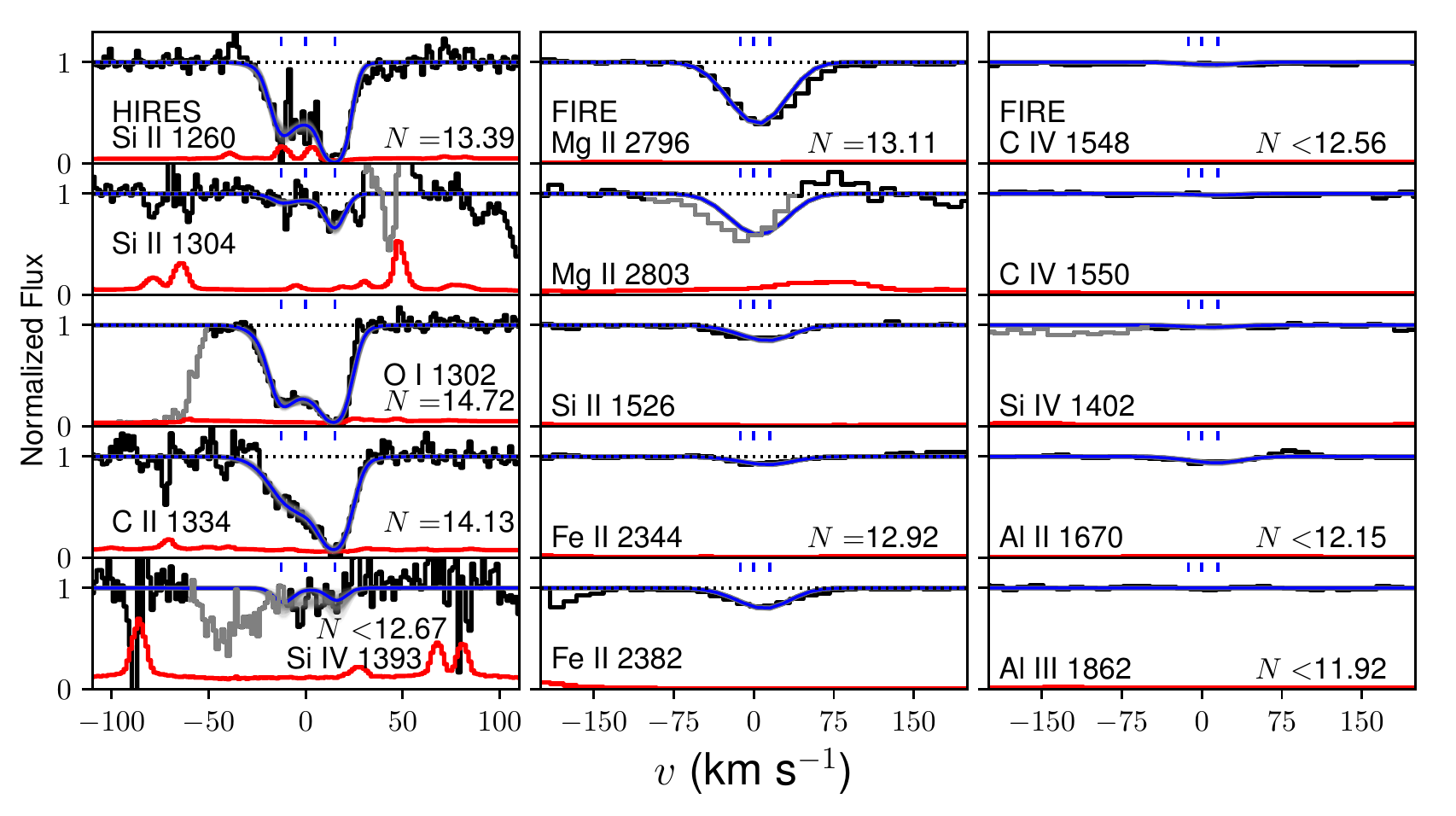}
\caption{This absorption system, at $z=6.1435$ along the line of sight to J0100+2802, is representative of those typically seen at $z\gtrsim6$. It exhibits absorption of low-ionization species, including \oi\ which suggests an appreciable fraction of neutral gas, but no \civ, despite higher SNR than most other FIRE spectra. See Figure \ref{fig:zabs6.18} for velocity plot description.}
\label{fig:zabs6.14}
\end{figure*}

\begin{figure*}
\includegraphics{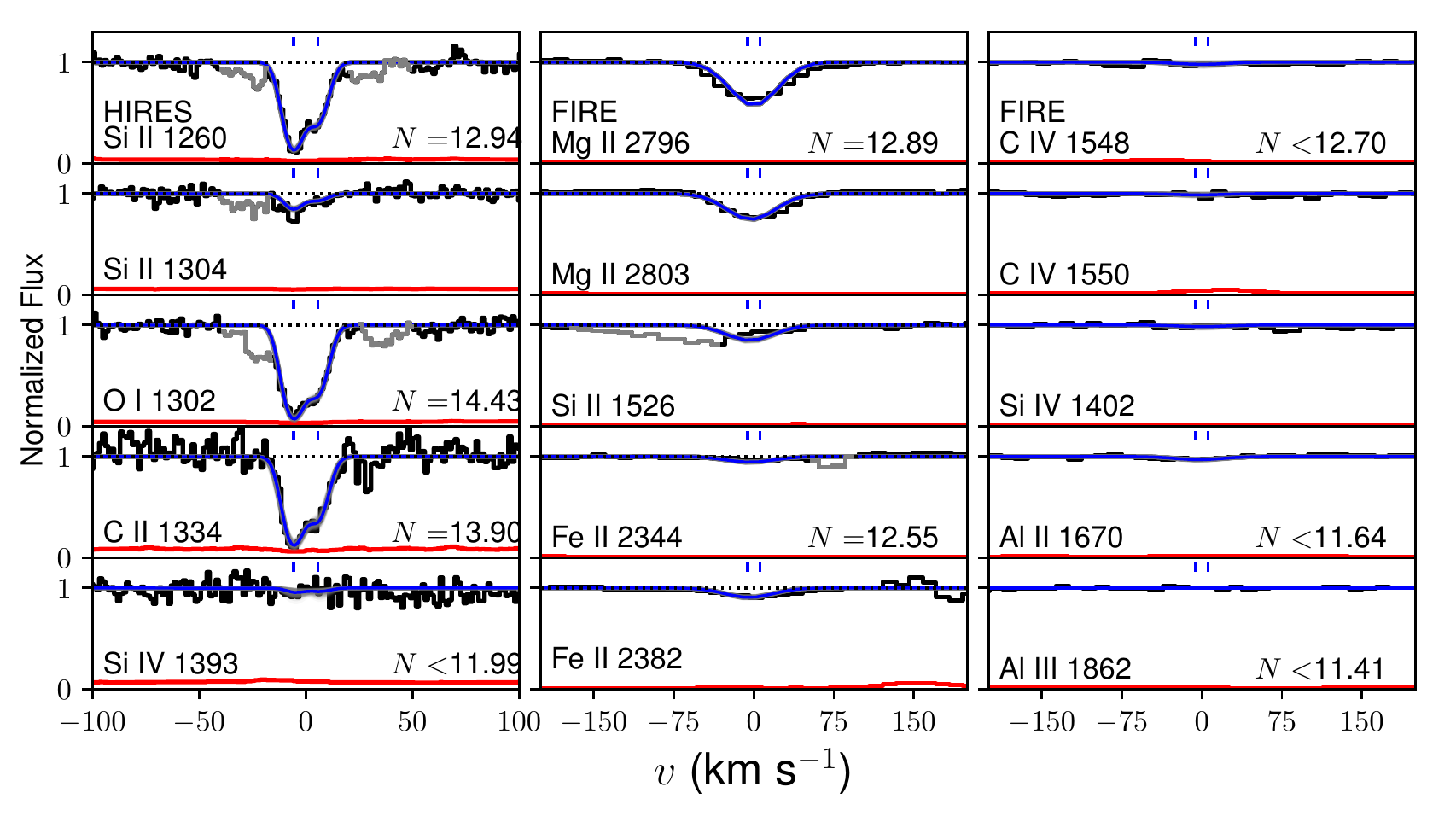}
\caption{This absorption system at $z=6.1117$ along the line of sight to J0100+2802 also has strong, narrow low-ionization absorption and no high-ionization absorption, even with minimal noise around the \siiv\ 1393\AA\ transition covered by HIRES. See Figure \ref{fig:zabs6.18} for velocity plot description.}
\label{fig:zabs6.11}
\end{figure*}

Our deep FIRE and HIRES observations of the ultraluminous QSO SDSS J0100+2802 revealed seven foreground absorbers, of which three lie at $z>6$ and are therefore ideal for detailed study at high resolution. HIRES and FIRE spectra of these absorbers are plotted in Figures \ref{fig:zabs6.18} through \ref{fig:zabs4.8}. Total column densities for the seven systems with \civ\ redward of the Ly$\alpha$ forest (i.e., $z\gtrsim4.73$) are given in Table \ref{tab:J0100abs}; component-by-component measurements are in Table \ref{tab:J0100abs2}.

The trend of lower-ionization toward higher redshift, noted above for the 50-object sample, is echoed even within the heavy-element absorbers measured in this single spectrum. The three systems of lowest redshift ($z=4.86,5.11,5.34$) all exhibit strong \civ\ in conjunction with low-ionization lines, while the next three ($z=5.79,6.11,6.14$) are either undetected in \civ\ despite strong low-ionization absorption or have considerably less \civ\ than \cii. The only exception is at $z=6.19$, which is only seen in \civ\ and may be affected by proximity to the background QSO, which is separated by only $\sim5900$ \kms.

\subsubsection{Velocity Structure}

We find simple and narrow kinematics in the high-redshift, low-ionization absorber population, similar to prior studies \citep{2017MNRAS.470.1919B,2006ApJ...640...69B}. The Doppler parameters for individual components from our MCMC Voigt profile fits are often only marginally resolved even by HIRES at $b=6-7$ km s$^{-1}$, and jointly fitting to multiple low-ionization species suggests that temperatures of $\lesssim10^3$ K are plausible for some of these absorbers, colder than is typically seen in the lower redshift circumgalactic medium. Moreover the velocity spread between centroids of various subcomponents spans a small range $\Delta v \lesssim 50$ \kms.

The high-ionization absorbers ($z=6.19,5.34, 5.11, 4.86$) span larger velocity ranges, generally larger than 100 \kms, with the largest spanning over 300 \kms. Similarly, the discrete absorption components are more widely separated in velocity.

\subsubsection{J0100+2802 Absorption System: $z=6.187$}

This system (Figure \ref{fig:zabs6.18}) is detected only in \civ\ and \siiv, and has two distinct yet blended components in the FIRE spectrum, with intrinsic width $\Delta v\approx 250$ \kms\ and spacing of $\Delta v\approx 125$ \kms. This combination presents rarely at $z>6$ and (as noted earlier) may be affected by proximity to the background quasar at $z=6.33$ \citep{2016ApJ...830...53W}.

The HIRES data reveal a very weak ($\log N=11.59$) \siii\ 1260 absorption feature aligned with the peak of the \civ\ absorption. This component is very narrow; fitting yields a Doppler parameter of $b\approx2$ \kms, implying that the line is unresolved. Careful checks of individual exposures, sky-line residuals, and telluric correction spectra suggest this feature is indeed real; if so then this system would have $\X{C}<0.04$ and $\X{Si}=0.07$.

Since the incidence rate of intervening high-ionization absorption systems is small at $z\gtrsim6$, we considered the possibility that this absorption system is a high-velocity outflow associated with the QSO. If this absorber is associated with the quasar, it would have a velocity of $v_{\text{abs}}\sim5900$ \kms\ relative to the quasar. \citet{2016MNRAS.462.3285P} find a covering fraction of \civ\ absorbers with $W_{1548}>0.2$\AA\ within 5000 \kms\ of $z=3.5-4.5$ quasars of $f_{1548}\approx0.4$, an excess of 0.2 over the covering fraction of intervening \civ\ absorbers at the same redshift (this system has $W_{1548}=0.26\pm0.01\rm{\AA}$). The ratio of intervening absorbers to intrinsic absorbers decreases with increasing redshift. They also note that \ion{N}{5} is a more definitive tracer of intrinsic absorbers within 5000 \kms\ (although it decreases in covering fraction by a factor of 2 from between 2500 to 5000 \kms), while we find an upper limit of $\log\N{NV}<13.0$. Hence, it is inconclusive if this absorber is intrinsic to the QSO or intervening. Given the detection of several other intervening \civ\ absorbers at $z>6$ \citep[e.g.][]{2017MNRAS.470.1919B}, we assume this to be an intervening absorber in the rest of this work.

\subsubsection{J0100+2802 Absorption System: $z=6.1435$}

This system (Figure \ref{fig:zabs6.14}) features three distinct components of low-ionization gas, within a narrow velocity envelope of $\Delta v\sim60$ \kms. The strongest component is mildly saturated in \cii\ 1334, \oi\ 1302, and \siii\ 1260, but the weaker \siii\ 1304 line allows for a reliable measurement of \N{Si II}. We measure very sensitive upper limits on allowed high-ionization absorption, resulting in $\X{C}>37.2$ and $\X{Si}>13.2$. The two stronger components yield Doppler parameters ranging from $b=6-7~\kms$ while the weaker central component is somewhat wider. A model with only two components yields similar total column density but slightly lower likelihood.

Because these lines are only very mildly saturated, the MCMC posterior total column density distributions extend just $\lesssim0.1$ dex above the median, so we represent these as measurements and not lower limits in the remaining analysis.

\subsubsection{J0100+2802 Absorption System: $z=6.1117$}

This is qualitatively similar to the previous system, with two slightly weaker and narrower low-ionization components (Figure \ref{fig:zabs6.11}). The Voigt profile fits suggest the absorbing gas could have temperatures as low as $100\mathrm{~K}$, although degeneracy between turbulent and thermal broadening allows for temperatures of several times $10^3\rm{~K}$.

\siiv\ 1393\AA\ falls on a clean part of the HIRES spectrum but is not detected, yielding a sensitive upper limit of $\log \N{SiIV}<11.99$, or $\X{Si}>8.9$. The carbon ratio $\X{C}>15.8$, is large but slightly less than the previous system on account of a lower $\N{CII}$.

While there is some absorption present on the wings of the \siii\ and \oi\ profiles, it is not kinematically aligned, and can also be explained by lower-redshift interlopers. We do not include them in the fit, and in any case doing so would only lead to a small fractional increase in total column density.

\begin{figure*} 
\includegraphics{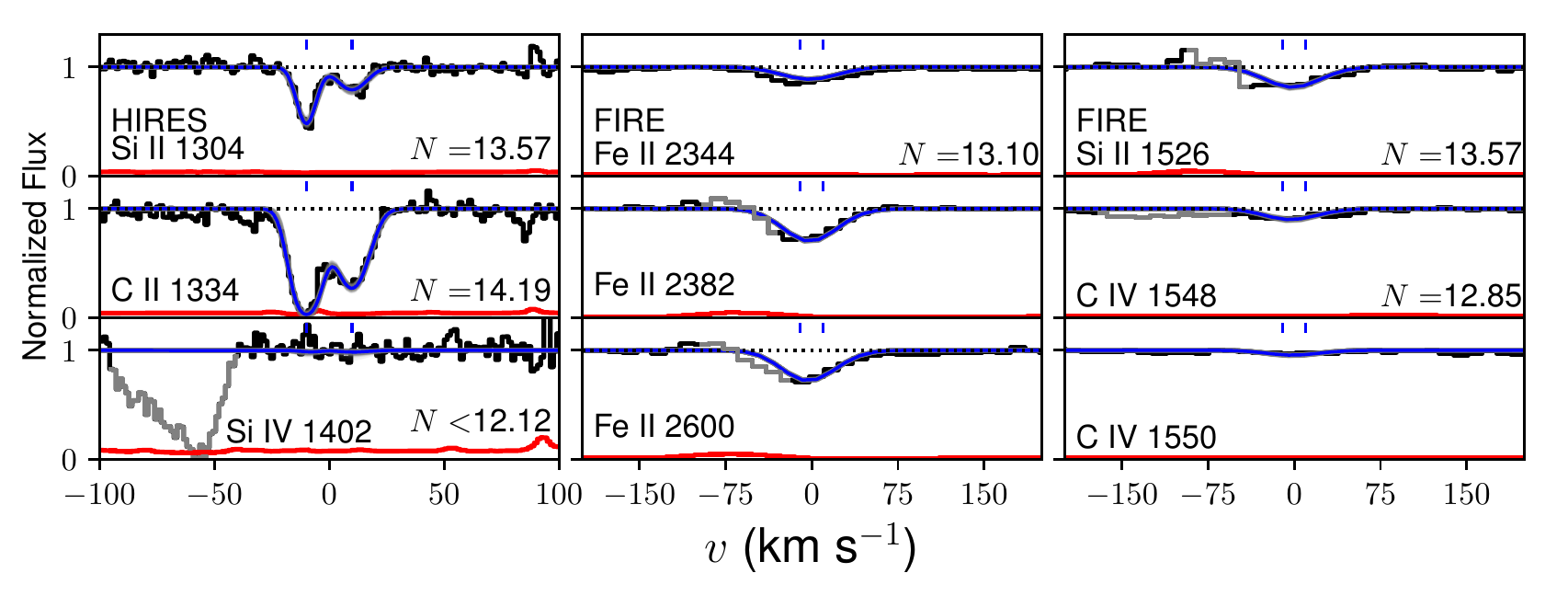}
\caption{This absorption system at $z=5.7979$ along the line of sight to J0100+2800 is much like the other high-redshift low-ionization absorbers. The absorption blueward of \civ\ 1548\AA\ is a known \mgii\ interloper, and several transitions are blended with incompletely-corrected telluric absorption evident in the error array. See Figure \ref{fig:zabs6.18} for velocity plot description.}
\label{fig:zabs5.79}
\end{figure*}

\begin{figure*}
\includegraphics{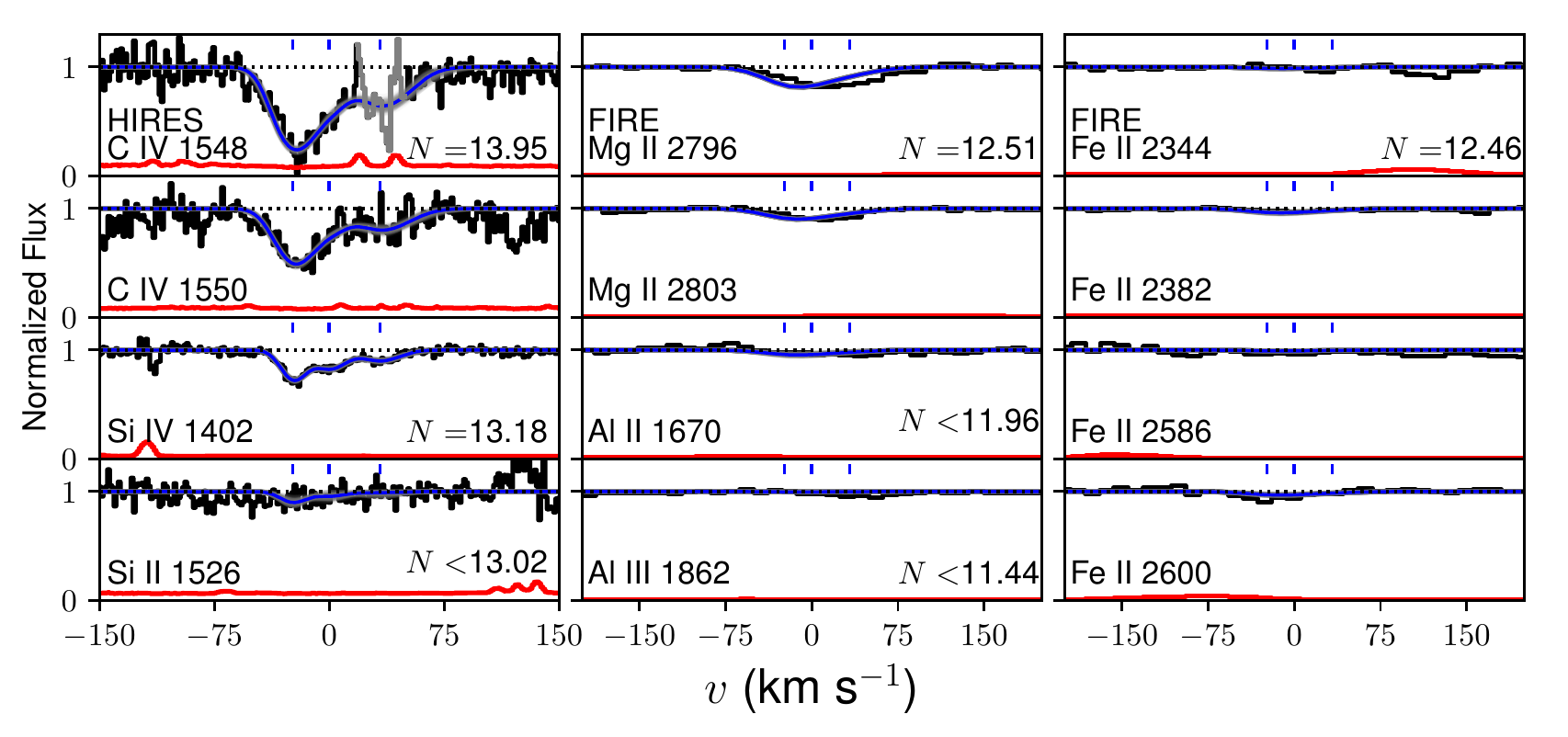}
\caption{This absorption system at $z=5.3383$ along the line of sight to J0100+2802, has \civ\ absorption resolved by HIRES and weaker low-ionization absorption. See Figure \ref{fig:zabs6.18} for velocity plot description.}
\label{fig:zabs5.33}
\end{figure*}

\begin{figure*}
\includegraphics{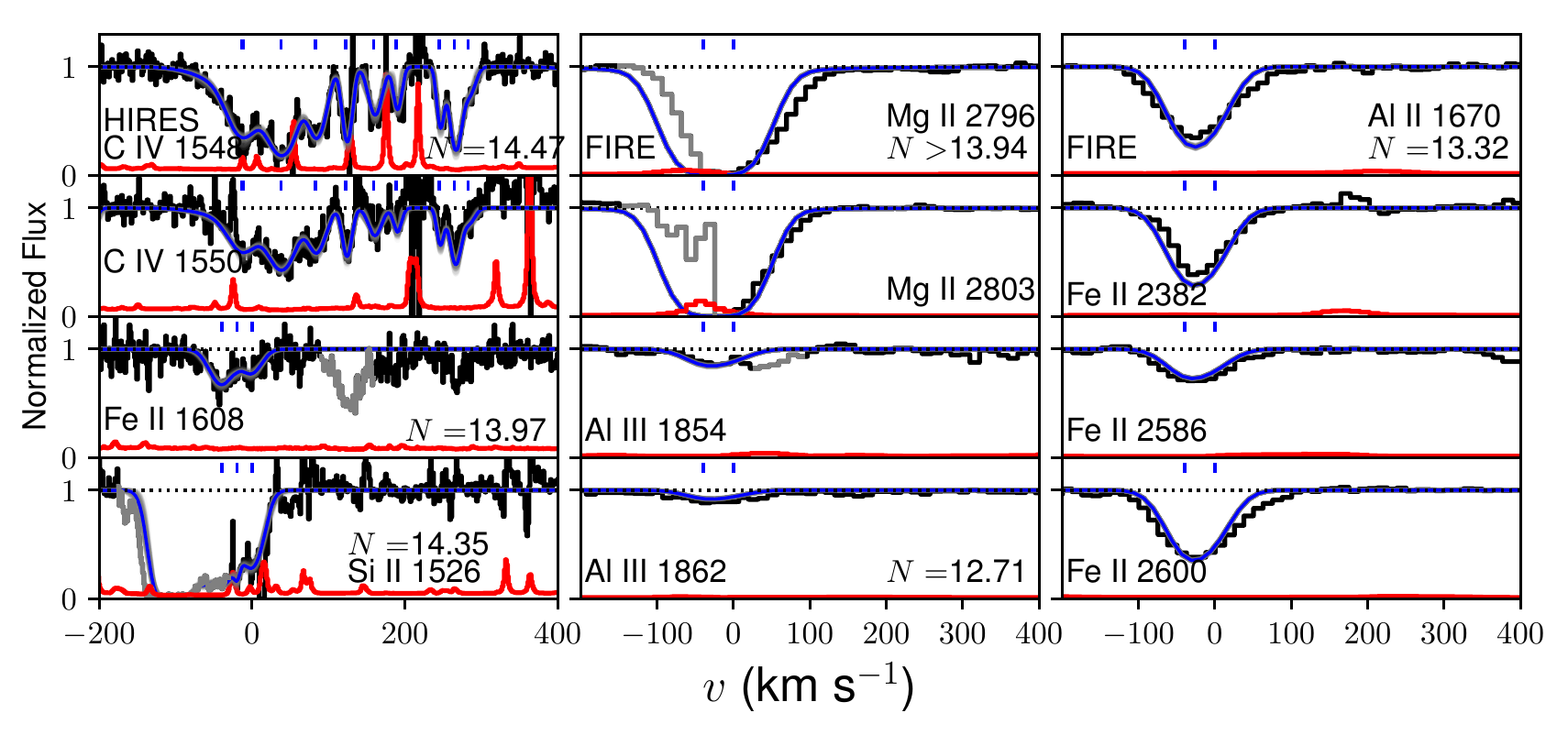}
\caption{The absorption system at $z=5.11$ along the line of sight to J0100+2802 has low-ionization absorption is associated with a portion of the \civ\ profile; most of the \civ\ does not have corresponding low-ionization absorption. The saturated \mgii\ profiles are made asymmetric by telluric corrections. See Figure \ref{fig:zabs6.18} for velocity plot description.}
\label{fig:zabs5.11}
\end{figure*}

\subsubsection{J0100+2802 Absorption System: $z=5.7979$}

Like the previous two low-ionization systems, this absorber (Figure \ref{fig:zabs5.79}) has a narrow, two-component profile of total width ($\Delta v\sim50$ \kms) in singly-ionized species. However, unlike the higher redshift absorbers, it appears to have weak \civ\ absorption, with $\log\N{CIV}=12.85$ (and $W_r=0.035\pm0.005$, making this a 7$\sigma$ detection), for an ionization fraction ratio of $\X{C}=22$ (there is interloping $z=2.75$ \mgii\ absorption blueward of \civ\ 1548\AA\ and no detection at 1550\AA; we treat this as detected \civ\ since absorption at 1550\AA\ is clear an X-Shooter spectrum (G. Becker, private communication)). The \siiv\ 1393 transition is similarly blended with \civ\ absorption at $z=5.1083$, but the undetected 1402\AA\ line in our HIRES spectrum implies $\N{Si IV}<12.12$, and a large ratio of $\X{Si}>28.2$. \citet{2019MNRAS.483...19M} find $\log\N{CIV}=13.1$ for this system; even with this larger high ionization gas measurement, the ionization ratio of $\X{C}=12$ still favors the neutral gas.

\subsubsection{J0100+2802 Absorption System: $z=5.3383$}
Excepting the possible quasar-intrinsic system at $z=6.187$, this is the highest redshift system in J0100+2802 with stronger triply-ionized carbon absorption. Unlike the unusual $z=6.187$ system, this absorber is relatively typical for its redshift. The \civ\ and \siiv\ profiles span $\Delta v\sim 100$ \kms\ and have three components, as is most evident in the high signal-to-noise \siiv\ 1402\AA\ line. The FIRE spectrum does not resolve this kinematic substructure, but our forced Voigt profile fits (with three components of $z$ and $b$ fixed to the HIRES values) yield $\X{Si}<0.7$.  \cii\ for this system falls within the Ly-$\alpha$ forest and is therefore inaccessible, but using Equation 1 to convert from the \mgii\ proxy we would find $\log\N{CII}=13.34$, or $\X{C}<0.24$.

\subsubsection{J0100+2802 Absorption System: $z=5.1083$}
The strongest $z>5$ absorption system in the J0100 spectrum (Figure \ref{fig:zabs5.11}) has a complex multi-component \civ\ profile spread over $\sim400$ \kms. There is narrow ($\Delta v\sim 50~\kms$) and very strong absorption from \cii, \feii, \alii, and \siii, aligned with the bluest portion of the \civ\ profile. Most of the \civ\ profile has no corresponding singly-ionized absorption, but where the singly ionized species do appear they are heavily saturated implying high column density. An intervening \mgii\ absorber at $z=2.3$ partially obscures the \siii\ 1526\AA\ line, but is jointly fit with its 2796\AA\ line to estimate $\N{SiII}$.

\begin{figure*}[t]
\includegraphics{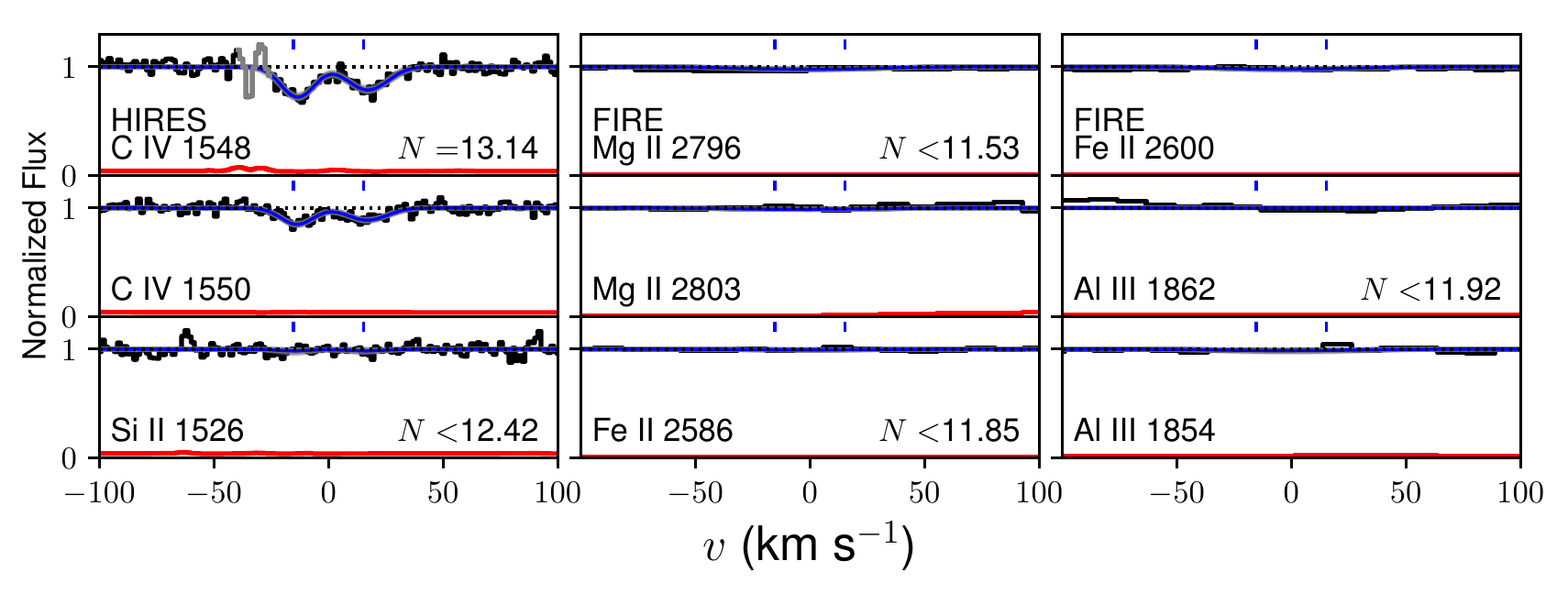}
\caption{The absorption system at $z=4.8751$ along the line of sight to J0100+2802 is included because it has \civ\ coverage. See Figure \ref{fig:zabs6.18} for velocity plot description.}
\label{fig:zabs4.8}
\end{figure*}

The changing value of $\X{C}$ across the profile (assuming a large value of $\N{CII}$ expected given such strong \mgii) is typical of DLAs at lower redshift. It requires variable levels of ionization arising from distinct gas phases, such as cold neutral clumps embedded in a lower-density ionized circumgalactic medium. 

\subsubsection{J0100+2802 Absorption System: $z=4.8751$}

This $z<5$ absorber has two weak \civ\ components spanning $\Delta v\sim100$ \kms, and no corresponding singly-ionized species (Figure \ref{fig:zabs4.8}). This is the most common type of metal-line absorber at $z<5$ \citep[e.g.,][]{1998AJ....115.2184S}, yet is all but absent at $z>6$. Again using our upper limit on \mgii\ as a proxy for \cii , we find $\X{C}<0.17$.

\section{Discussion}\label{sec:interp}

\subsection{Low-Ionization Absorbers as Analogs of Metal-Poor DLAs}

In Section \ref{sec:ratios} and Figure \ref{fig:highz_ratios} we argued that individual low-ionization metal absorbers---which dominate the $z>5.7$ population---have ratios of low- to high-ionization carbon resembling the neutral phase for lower redshift DLAs. Consistent with \citet{2011ApJ...735...93B}, we also find that ratios of low-ionization species alone (\oi, \cii, \siii) are similar to those of metal-poor DLAs and Sub-DLAs. Although we cannot measure \nhi\ at $z\sim 6$, the heavy element column densities would yield heavy element abundances in the metal-poor regime if one assumes typical DLA values. In this section we explore whether statistics of the population are collectively consistent with an extrapolation of metallicities and incidence rates for low-redshift DLAs and/or LLS, and discuss notable differences.

In the discussion below we use the canonical \hi-based definitions
for different absorber classes at lower redshift:
\begin{itemize}
\item{{\bf Lyman-limit systems (LLSs)} have $\log \N{HI}\ge17.2$. They are optically thick to Lyman continuum photons but significantly ionized at lower redshift, with typical ionization parameters of $-2<\log U<-3$. Their median abundance at $z\sim 3-4$ is $\mathrm{[X/H]}\sim-2.5$ with 0.5 dex scatter \citep{2015ApJ...812...58C,2016ApJ...833..270G,2016MNRAS.455.4100F}, and they exhibit both low- and high-ionization species of carbon and silicon at these redshifts.}
\item{ {\bf Damped Lyman-$\bm{\alpha}$ Systems (DLAs)} have $\log\N{HI}\ge20.3$ and are usually considered fully self-shielded and neutral ($\log U\lesssim -4$)---especially where \oi\ is detected---so ionization corrections are not used in calculating abundances. At $z\sim 3.5$ their median abundance (corrected for dust depletion) of $\mathrm{[Fe/H]}\sim-1.5\pm0.5$ \citep[e.g.,][]{2018A&A...611A..76D} is markedly higher than that of LLSs. At these levels, the transitions discussed in this paper are heavily saturated.}
\item{{\bf Sub-DLAs} (sometimes called Super-LLS) are intermediate between the neutral DLAs and ionized LLS at $19.0<\log\N{HI}<20.3$. They therefore require slight ionization corrections to derive heavy-element abundances. The median sub-DLA abundance slightly exceeds that of DLAs at $z\lesssim 4$ \citep{2016MNRAS.458.4074Q} but extrapolation of its redshift evolution suggests that sub-DLA abundances may cross below DLAs at $z>4$.}
\end{itemize}

To avoid confusion, since DLAs, Sub-DLAs and LLS are defined by $\N{HI}$ we instead refer to the \cii-dominant high-redshift systems as \textbf{Low-Ionization Absorbers (LIAs)} and reserve the \hi-based nomenclature for lower-redshift comparison samples.

\subsubsection{Incidence Rate of LIAs Versus DLA Extrapolation}\label{sec:incidence}

First, we consider how the number of detected $z\sim 6$ LIAs compares with predictions made by extrapolating the comoving absorber density of DLAs at lower redshift. The absorber density is estimated from the number of detections $\mathcal{N}$:
\begin{equation}
\dNdX=\frac{\mathcal{N}}{\Delta X}
\end{equation} 
where the unitless distance $\Delta X(z)$ resembles a redshift interval $\Delta z$ but is corrected for cosmic expansion, such that an unchanging population with constant comoving density and cross section exhibits constant $\dNdXt$ \citep{1969ApJ...156L...7B}. There are $\mathcal{N}=17$ absorption systems detected in \cii\ or \mgii\ in our search at $z\ge5.7$ (33 with $z\ge5.0$). The pathlength searched is $\Delta X=99.0$ (218.1); for most of the pathlength at $z<5.7$ in our FIRE data, \cii\ is inaccessible. Assuming that our search identified all absorption systems, this yields an LIA number density of $\frac{d\mathcal{N}}{dX}$=0.17 (0.15), consistent with the findings of \citet{2011ApJ...735...93B}; the typical detection threshold is $\N{CII}\sim13.5$ cm$^{-2}$ although this varies across and within sightlines. While 100\% completeness is implausible, the detection rate is likely close to unity for the stronger absorbers, at $\N{CII}\gtrsim14.0$.  An important note is that LIAs weaker than those observed in our data are presumed to exist, and in fact weaker \mgii\ absorbers evolve differently from stronger ones at lower redshifts, implying they may represent a physically distinct population \citep{2012ApJ...761..112M,2017ApJ...850..188C,2017arXiv170105624M}. Hence, the following discussion regarding the nature of LIAs is limited to the class of absorbers detectable in our data, as weaker absorption systems may have different ionization fractions. Similarly, the nondetection of high ionization gas in LIAs is sensitivity dependent, although our limits already constrain absorbers' ionization fractions less commonly seen at lower redshift.

The DLA and sub-DLA incidence rates may be extrapolated from low redshift using the \hi\ column density distribution function $f_\mathrm{HI}(N,X)=\frac{\partial^2\mathcal{N}}{\partial N\partial X}$ which has been fit by either double power-laws or $\Gamma$-functions:
\begin{eqnarray}
f_\mathrm{HI}(N,X)&=&
\begin{cases}
k_d(N/N_d)^{\alpha_1} & N<N_d \\
 k_d(N/N_d)^{\alpha_2} & N>N_d 
 \end{cases}\\
 f_\mathrm{HI}(N,X)&=&k_g\left(N/N_g\right)^{\alpha_g}e^{(-N/N_g)}.
 \end{eqnarray}
\citet{2009ApJ...696.1543P} find that the shape of the distribution is invariant for $z=2-4$. Integrating over \nhi\ provides $\dNdXt$. For recommended power-law indices $-3<\alpha<-1$, $f_\mathrm{HI}(N,X)$ increases rapidly as the lower integration limit decreases. Here we consider how far one must reduce this limit of integration to match the incidence rate of LIAs. Because the power law is steep (or exponentially declining) at high $\N{HI}$ the integral is insensitive to the upper bound.

We used several \hi\ distribution functions from the literature, including estimates for DLAs at $4.0<z<5.5$ \citep{2009ApJ...696.1543P}, DLAs at $2.0<z<4.0$ \citep{2009A&A...505.1087N}, and DLAs at $3.2<z<5.3$ \citep{2016MNRAS.456.4488S}. For $\N{HI}>20.3$, these distribution functions all yield predictions of $\frac{dN}{dX}\sim0.10$ (consistent with other estimates of DLA incidence rates at high redshift, e.g., \citealp{2015MNRAS.452..217C}). We observed nearly twice as many LIAs per unit pathlength, not even correcting for incompleteness, suggesting that the LIA population may not consist entirely of DLA analogs.

\begin{figure}[]
\includegraphics[width=\linewidth]{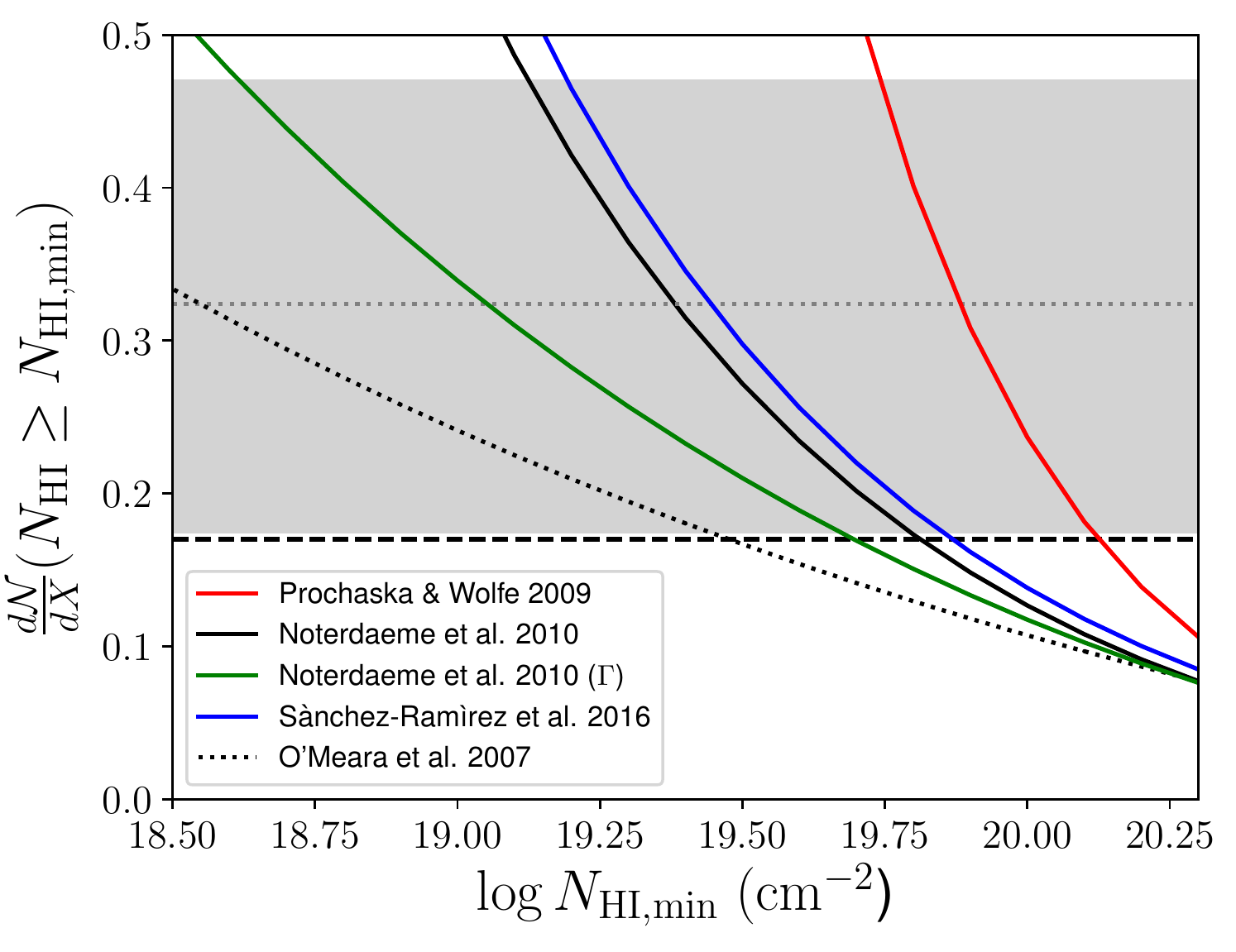}
\caption{Estimated incidence rate (\dNdXt) of predominantly neutral absorption systems, as a function of the lowest \hi\ column density included, based on \hi\ column density distribution functions at $z=3$ to 5. The solid colored lines use distribution functions based on DLAs, and the dotted line uses the measured sub-DLA distribution and same DLA distribution function as the green line. The horizontal line indicates our observed LIA incidence rate at $z>5.7$ of 0.17 (with a loosely-defined detection threshold of $\N{CII}>13.5$ cm$^{-2})$}. This demonstrates that we do not require low-ionization absorption from a population of LLSs to explain the incidence rate of LIAs at high redshift in our sample. To illustrate a similar completeness-corrected measure, the gray dotted horizontal line and shaded region denote the measurement and 1$-\sigma$ confidence interval of \dNdXt\ for \mgii\ absorbers with $W_{r,2796}>0.3$\AA\ from \citep{2017ApJ...850..188C}.
\label{fig:DLA_lx}
\end{figure}

However Figure \ref{fig:DLA_lx} illustrates the rapid increase in $\dNdXt$ as the lower \hi\ limit of integration is reduced. The slope of $f_\mathrm{HI}(N,X)$ is relatively uncertain in this column density range on account of the difficulty in measuring \nhi\ when absorption is saturated in the line core but does not produce strong damping wings \citep{2007ApJ...656..666O,2009ApJ...705L.113P}. Yet despite the choice of slope (indicated for different surveys by color), one only needs to generate heavy-element absorption in systems with $\log\N{HI}\gtrsim 19.8$ to reproduce the observed number density of LIAs (dashed horizontal line). This is only a factor of three lower in \nhi\ than the canonical DLA limit.

Viewed another way, the extrapolated number density of LLSs at $z>5.7$ is much larger than the observed rate of LIAs, not even accounting for the overall change in \hi\ ionization within the cosmic volume at $z>6$. LLSs outnumber DLAs by a factor of $\sim10$ at $z=4.4$  and both LLSs incidence rate and the ratio of LLSs to DLAs increase between $z=3.5$ and 4.4 \citep{2010ApJ...718..392P}. Supposing our metal-line survey is only 50$\%$ complete, the corresponding $2\times$ increase in the number density of low ionization absorbers could still easily be explained by DLAs and sub-DLAs alone. We demonstrate this with the completeness-corrected \mgii\ incidence rate (gray dotted horizontal line) and $1-\sigma$ uncertainty at $z\sim6.2$ from \citet{2017ApJ...850..188C}. While the completeness corrections do increase the incidence rate by a factor of 2, it is still evident that if an appreciable fraction of LLSs yielded LIAs at $z\sim6$, we would find a significantly larger incidence rate. (Although the shallower slope of this relation at lower \N{HI} results in these models overpredicting the incidence rate.)   This conclusion is only strengthened if $\frac{dN}{dX}$ of DLAs continues to increase from $z=2$ to $z=5$, as has been suggested (\citealp{2015MNRAS.452..217C,2009ApJ...696.1543P}; c.f. \citealp{2010ApJ...721.1448S}).

If all LIAs are DLAs and sub-DLAs, it is likely that currently-observable metal lines are tracing early precursors of the interstellar and/or circumgalactic medium, and do not yet represent truly intergalactic matter.  At $z\sim3$ a large fraction of discrete \cii\ and \civ\ absorbers are associated with LLSs and even weaker \hi\ systems, and many of these \hi\ absorbers are presumed to reside in the outer circumgalactic halos of star-forming galaxies \citep{2012ApJ...750...67R,2014MNRAS.438..529R}. At $z>6$ these LLSs and weaker \hi\ absorbers are either too metal-poor for us to detect, or their heavy elements reside in ionization states that are inaccessible because of Lyman-$\alpha$ forest saturation or sensitivity constraints.

\subsubsection{Chemical Abundance Distributions for the DLA Hypothesis}

Supposing that LIAs of the class we observe (i.e., $\N{CII}\gtrsim13.5$) arise in high redshift analogs of neutral DLAs (or slightly ionized sub-DLAs), one may ask what metallicity would yield \cii, \siii\ or other heavy element column densities similar to those observed.

To test this, we take the compilation of $z=1-5$ ıDLA metallicities corrected for dust depletion \citep{2018A&A...611A..76D}, extrapolated to redshift six using their fit for evolution in the mean abundance, [Fe/H]$=-0.36-0.32z$, and redshift-invariant scatter of $\sigma=0.55$ dex. The authors argue that dust depletion requires negligible corrections to column densities at [O/H]$\lesssim-2$ \citep[see also][]{2016A&A...596A..97D}, and the derived abundances are broadly consistent with other surveys at $z\sim 5$ \citep{2018MNRAS.473.3559P,2016ApJ...830..158M,2012ApJ...755...89R,2014ApJ...782L..29R} and simulations \citep[e.g.,][]{2015MNRAS.453.3798M,2018MNRAS.476.4865R}. The resulting hypothetical metallicity distribution has a mean [Fe/H]$=-2.28$ and scatter of 0.55 dex.

We construct Monte Carlo distributions of $\N{CII}$ by drawing \hi\ absorbers randomly from the double power-law form of $f_{\rm HI}(N,X)$ \citep{2009A&A...505.1087N}. We then independently assign each absorber a random chemical abundance from a normal distribution with mean and standard deviation calculated as described above. We assume for this exercise that all carbon is in the \cii\ state, which should be very accurate for $\log \N{HI}>20.3$ (\citealp{2001ApJ...557.1007V} find ionization corrections of $<0.05$ dex), and also that relative abundances follow the Solar pattern \citep{2009ARA&A..47..481A}. This assumption simplifies the analysis but is not strictly true for sub-DLAs and LLS, as discussed below.

\begin{figure*}
\includegraphics[width=0.5\linewidth]{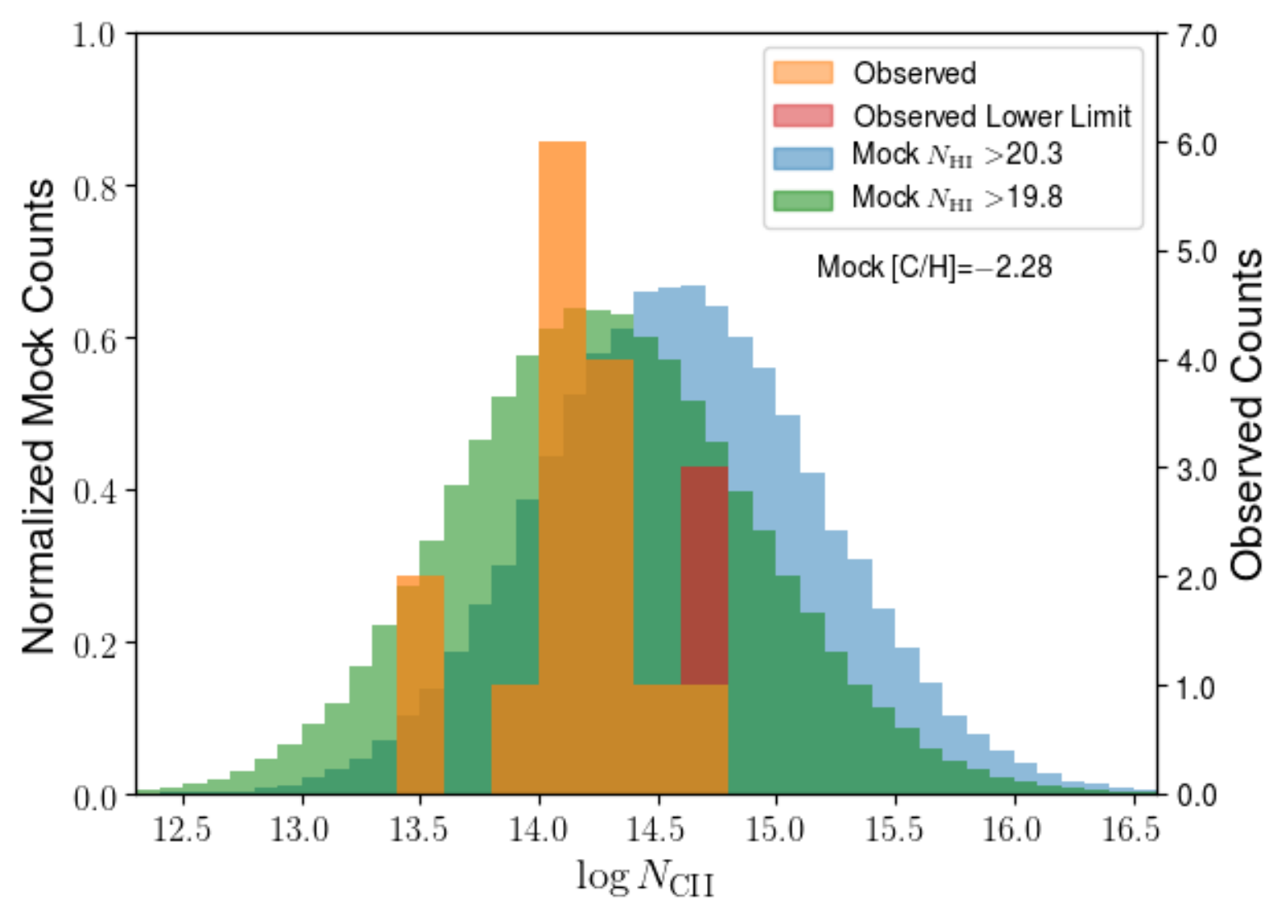}
\includegraphics[width=0.5\linewidth]{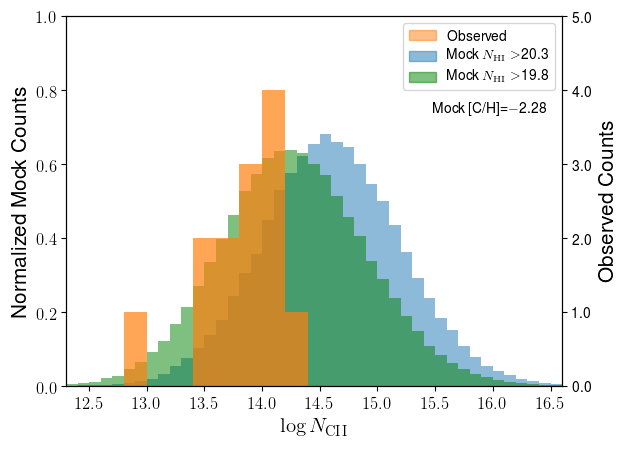}
\caption{Left: Mock \cii\ column densities, assuming a population of absorbers with $\langle\rm{[C/H]}\rangle=-2.28$ and standard deviation of 0.55, with 100\% of carbon in \cii, and an \N{HI} distribution function shape from \citet{2009A&A...505.1087N}. The blue histogram includes only canonical DLAs with $\log\N{HI}\ge20.3$, and the green extends down to $\log\N{HI}=19.8$. LIA \cii\ column densities observed from our FIRE sample (and QSO J0100+2802) are overlaid, with saturated lines indicated by red. The mock distributions are normalized, whereas the right-hand y-axis shows the counts for observations. Right: The observations now only consist of `high-fidelity' spectra, including QSO J0100+2802, SDSS J1148+5251 \citep{2006ApJ...640...69B} and ULAS J1120+0641 \citep{2017MNRAS.470.1919B}. While these spectra are more sensitive (e.g., the HIRES spectrum of J0100+2802 has a \cii\ nondetection at $\N{HI}<12.6$), no weaker \cii\ absorbers are detected.
\label{fig:mockN}}
\end{figure*}

Figure \ref{fig:mockN} shows the resultant $\N{CII}$ probability distributions, for absorbers with $\N{HI}>10^{20.3}$ cm$^{-2}$ (blue) and $\N{HI}>10^{19.8}$ cm$^{-2}$ (green). The latter limit of $\log\N{HI}=19.8$ cm$^{-2}$ was chosen because it also produced our best match between $\dNdXt$ from the \hi\ distribution and LIAs.  Histograms of observed \cii\ columns are overlaid in orange (with one lower limit in red). Integrating to smaller limits of $\N{HI}$ yields a lower mean $\N{CII}$ as expected. The mock distributions are not precisely normal (in $\log\N{CII}$) because a higher proportion of systems start with lower $\N{HI}$.

Evidently this extrapolation method approximately reproduces the observed mean \cii\ column density as well as its detection frequency. Notably, the predicted \cii, \siii, \feii, \mgii, and \oi\ profiles are unsaturated, which is highly unusual for DLAs at $z\sim 3$ which have a mean [O/H]$\sim -1.3$. In fact, the LIAs would be classified as metal-poor DLAs (MPDLAs, see Fig. 1) defined by [O/H]$< -2$. At $z\sim 3$ MPDLAs make up roughly 10\% of the full parent population (and even fewer sub-DLAs); for the distribution used in the Monte Carlo simulation 69\% of LIAs would meet the MPDLA criterion.

There is weak evidence that the LIA $\N{CII}$ distribution is narrower than the Monte Carlo result, which may be attributable to the small sample size. We do not consider the lack of systems at low column density significant, as our search may be highly incomplete in those regions. If one restricts attention to higher-SNR spectra of J0100+2801, ULAS1120+0641 \citep{2017MNRAS.470.1919B} and SDSS1148+5251 \citep{2011ApJ...735...93B} there is an increased proportion of systems at $13.5<\log \N{CII} < 14.0$ (Figure \ref{fig:mockN}), but only one at $\logN{CII}=12.5-13.5$ where the HIRES data should still be sensitive. The low end tail of the $\N{CII}$ distribution is mostly produced by absorbers with smaller values of $\log \N{HI}\sim 19.8$, so it is possible that ionization plays a role in suppressing these systems, although that interpretation is highly speculative. A two-sample Kolmogorov-Smirnov test comparing the full FIRE sample to this subsample suggests they do not have the same parent distribution ($P=0.01$), reflecting the likely incompleteness of the FIRE data at $13.5<\log \N{CII} < 14.0$.

The lack of high-$\N{CII}$ systems is more interesting, as our sample should be highly complete at $\log\N{CII}>14.75$, representing 20\% of the cumulative distribution for the green Monte Carlo simulation but only two of 17 LIAs. However we do not wish to overstate the significance of this discrepancy considering the many extrapolations required to generate the model.

Instead, the point of this heuristic exercise is to illustrate---even lacking measurements of $\N{HI}$---that the incidence rate and heavy element column densities of the LIAs may be reasonably explained as a high-redshift population of DLAs and sub-DLAs, where the large majority are metal-poor with [C/H]$<-2$. Absorbers of this demographic are very rare at moderate redshift, comprising at most $\sim 10\%$ of the DLA population at $z\sim 3$, even though {\it all} DLAs produce only 1/10 as many heavy-element absorbers as the LLS population at the same epoch.

\subsection{Systematic Search for LIA Analogs at Low-$z$\label{sec:analogs}}

\begin{figure*}
\centering
\includegraphics[width=0.24\linewidth]{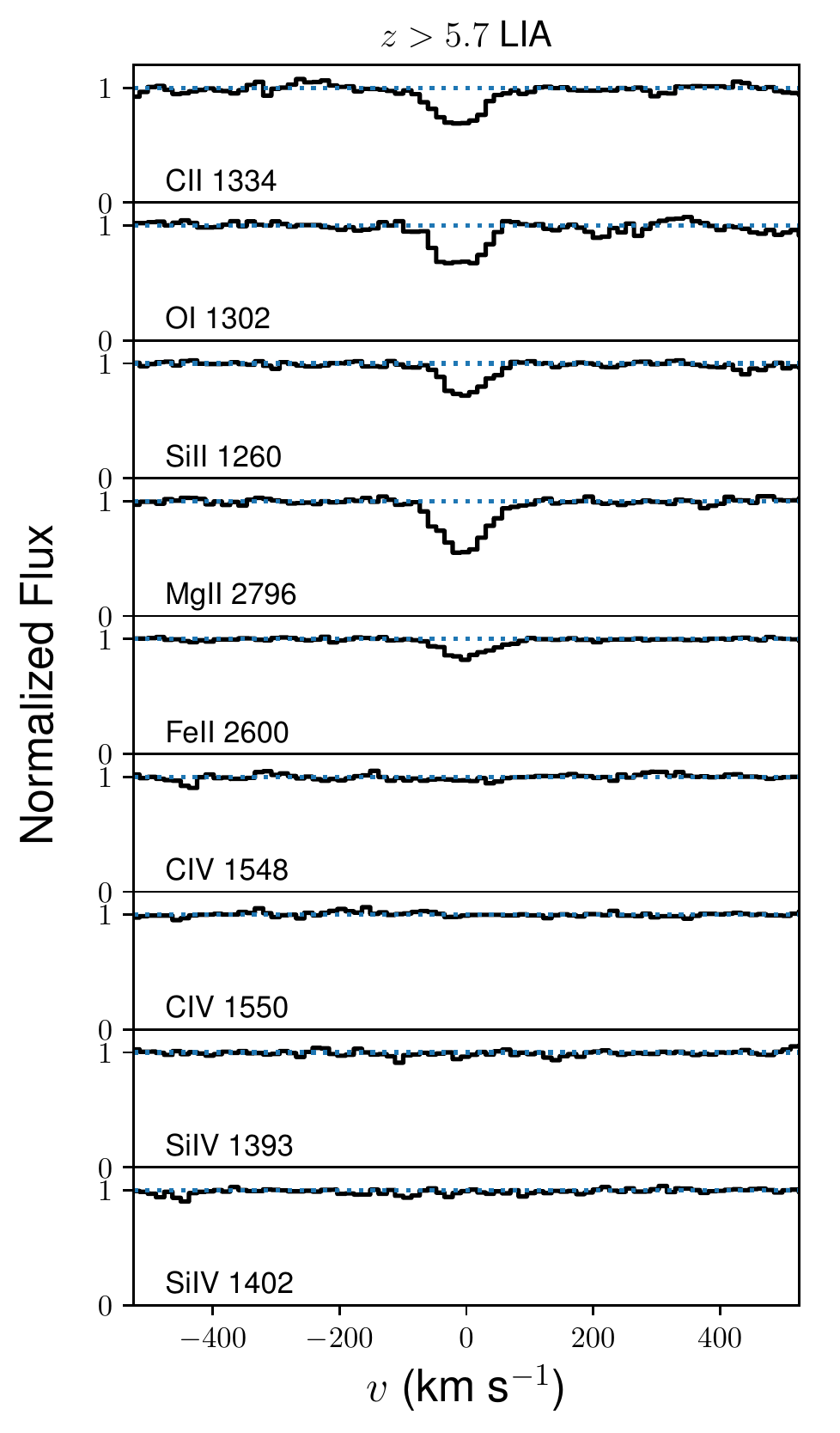}
\includegraphics[width=0.24\linewidth]{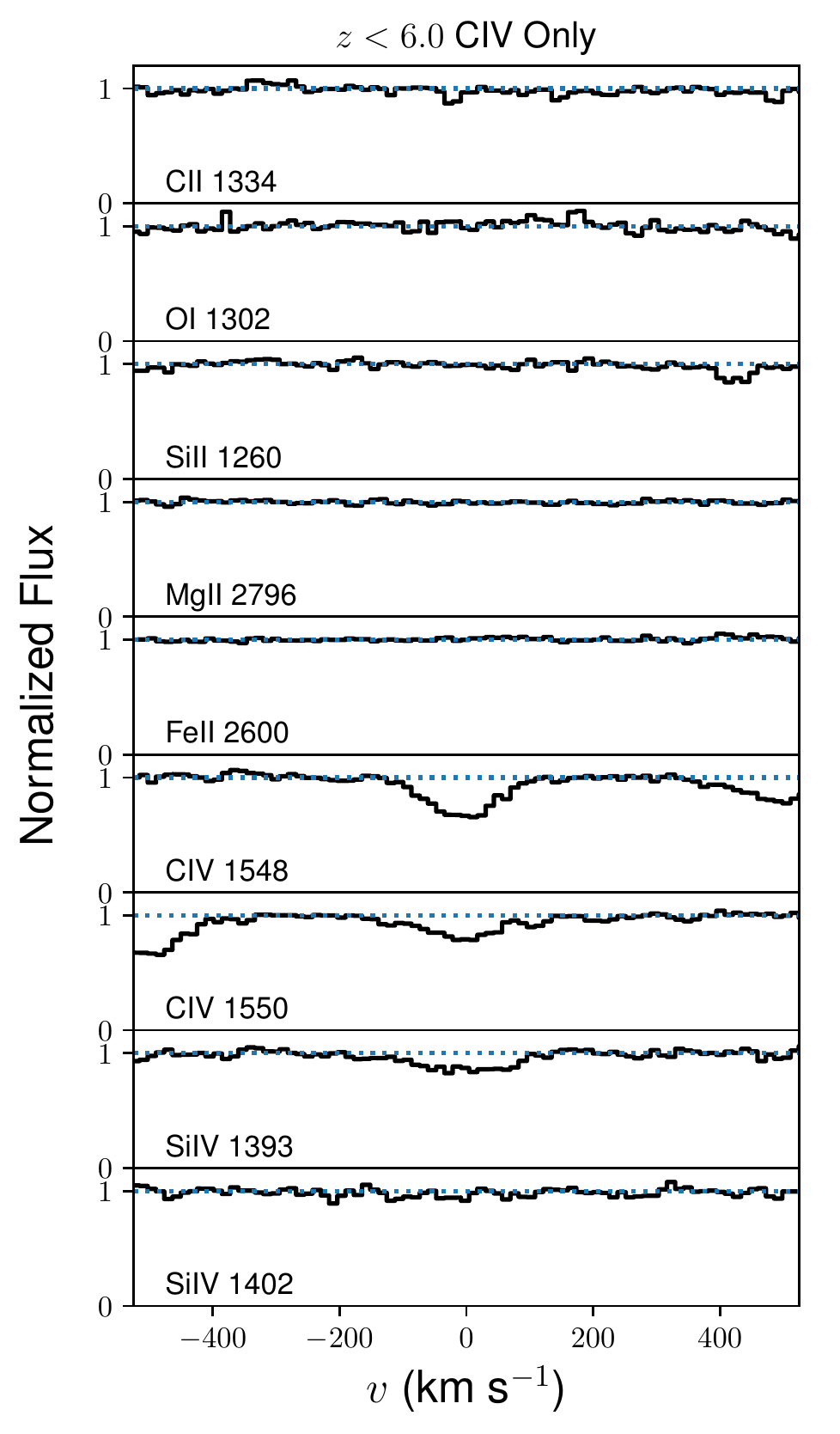}
\includegraphics[width=0.24\linewidth]{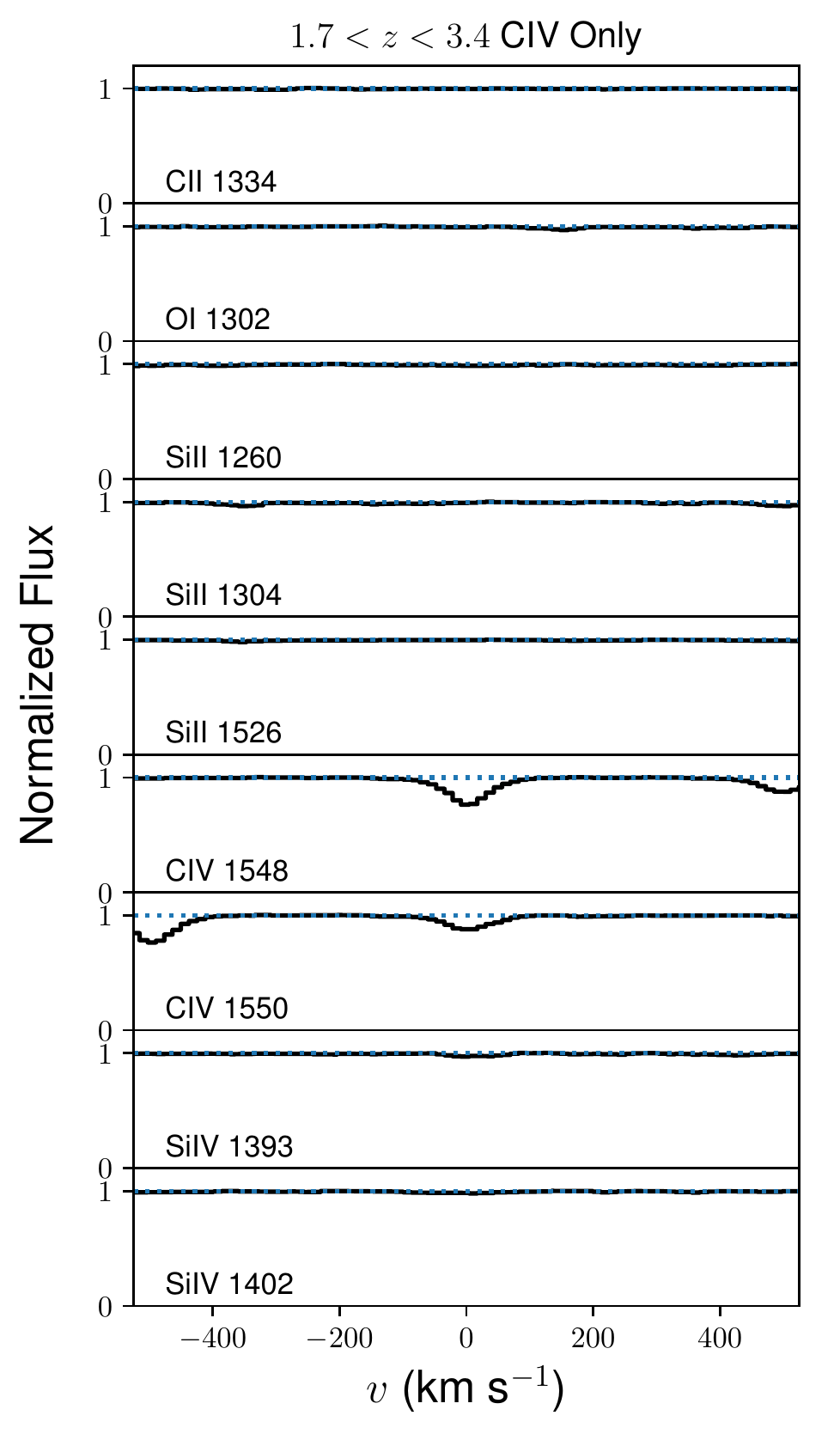}
\includegraphics[width=0.24\linewidth]{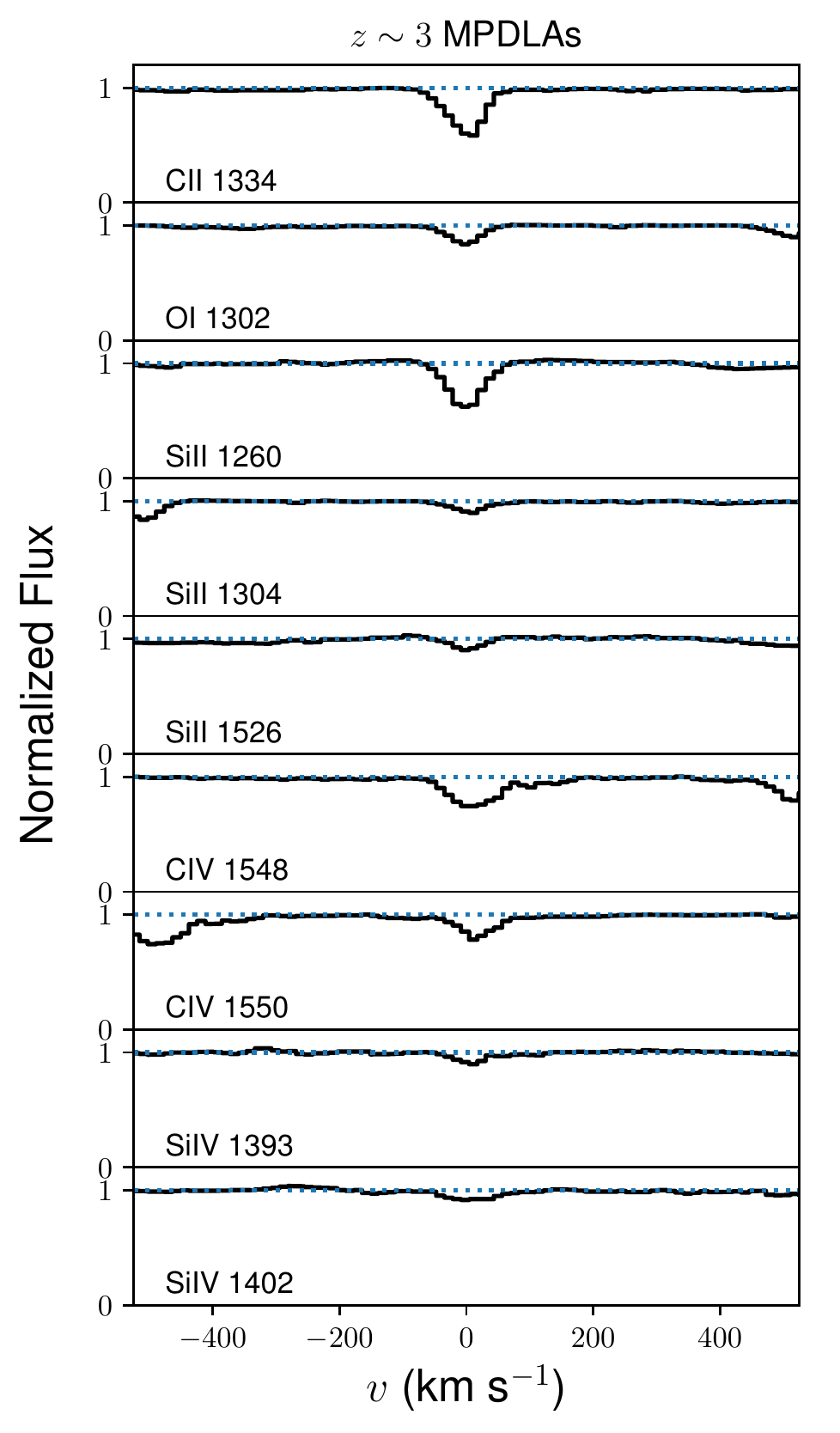}

\caption{Unweighted median stacks of (left) low-ionization absorbers at $z>5.7$ with no coincident \civ\ detection, (middle-left) \civ\ absorbers at $z<6$ with no coincident low-ionization detections, (middle-right) similar \civ\ absorbers at lower redshifit, and (right) metal-poor DLAs at $z\sim3$. The lower-redshift stacks are from HIRES data convolved with a Gaussian kernel of FWHM$=50$ \kms. Since HIRES data do not cover the \mgii\ or \feii\ multiplets, those panels are replaced with weaker \siii\ transitions. While stacking often reveals features below the SNR thresholds of individual spectra, no such features are apparent in either high-redshift stack, or the lower-redshift \civ. The MPDLAs, the closest lower-redshift analog we found to LIAs, have high-ionization gas as well.\label{fig:stack}}
\end{figure*}

LIAs are distinguished from $z<5$ metal absorbers by their lack of \civ\ and \siiv.  Figure \ref{fig:stack} (leftmost panel) shows an unweighted median stack of FIRE data from the full $z>5.7$ LIA sample, illustrating the presence of strong singly-ionized transitions (for species with first $E_{ion}<1$ Ryd) and non-detection of triply ionized species at higher SNR. The middle-left panel is a median stack of absorbers at $5<z<6$ \emph{without} low-ionization absorption detected, illustrating the difference between these types of absorption systems. Notably, in neither scenario did stacking reveal the presence of absorption from the undetected phase below the detection threshold of individual spectra. In lower redshift DLAs and LLS, \civ\ and \siiv\ are usually present at $\log\N{CIV}>13.5$ when $\log\nhi>17.3$ \citep{2015ApJS..221....2P}. 

We construct similar stacks from HIRES data (discussed below) of \civ\ absorbers without associated low-ions at $1.7<z<3.4$, and of $z\sim3$ MPDLAS. Much like the \civ\ absorbers at high-redshift, stacking does not reveal \cii\ or other low ions: in this sense, the \civ-only absorbers at high redshift have a qualitative analog at lower redshift. As shown in Figure \ref{fig:highz_ratios}, stronger \civ\ absorbers at lower redshift, not included in this stack, do have associated \cii. \ MPDLAs, which we find to be the best low-redshift analog for LIAs in the following discussion, clearly have appreciable amounts of gas present across different ionization states.

It is not straightforward to find low-redshift comparison samples to establish that \civ-free systems do not exist at later epochs but were missed by earlier surveys; because \civ\ is the predominant heavy-element ion at $2<z<5$ most surveys start by searching for this easily identified doublet and then measuring coincident \cii, rather than selecting blindly on the more easily misidentified \cii\ singlet and then measuring \civ\ strength as is required for LIAs. Many papers on \hi-selected DLAs only report column densities of singly-ionized species even though they exhibit \civ, because the \civ\ is ignored in abundance measurements on account of ionization uncertainty and likely arises from gas physically distinct to that bearing low-ionization species and the bulk of \hi.

To generate the low-redshift reference points in Figure \ref{fig:highz_ratios}, we used several samples. First, motivated by the common presence of \oi\ in our sample we examined purely \hi-selected DLAs surveyed with Keck-HIRES \citep{1999ApJS..121..369P,2001ApJS..137...21P,2003ApJS..147..227P,2007ApJS..171...29P}. Nearly all of these systems have heavily-saturated \cii, with lower limits at least 0.5 dex higher than our observed values at $z>5.7$.  However a small minority do have \civ\ non-detections.

Next we examined $z\sim 2-3$ absorbers explicitly preselected as candidate MPDLAs either from literature measurements at echellette \citep{2010ApJ...721....1P} and echelle resolution. The latter studies do not report $\N{CIV}$ so we downloaded archival spectra of the relevant QSOs from the HIRES and UVES archives to produce the measured \civ\ values (Table \ref{tab:MPDLAs}), and these spectra were used to produce the stack in Figure \ref{fig:stack}.  For studies published earlier than 2011, we use reference [O/H] values reported in \citet{2011MNRAS.417.1534C}, finding a median [O/H]=$-2.32$, and median $\N{CIV}=13.69$. 

\begin{deluxetable}{llllll}
\tabletypesize{\footnotesize}
\tablecaption{Metal-poor DLAs\label{tab:MPDLAs}}
\tablehead{\colhead{Quasar Name} & \colhead{z} & \colhead{\logN{CII}} & \colhead{\logN{CIV}} & \colhead{[O/H]} & \colhead{Ref}}
\startdata
J0035-0918      & 2.340 &    13.79 & $<$13.06 & -2.28 & 1\\
J0140-0839      & 3.696 &    13.88 &    12.65 & -2.75 & 2\\
J0307-4945      & 4.466 & $>$14.30 &    13.40 & -1.45 & 3\\
J0831+3358      & 2.304 & $>$14.33 & $<$13.41 & -2.01 & 4\\
Q0913+072       & 2.617 &    14.21 &    14.13 & -2.40 & 5\\
J1001+0343      & 3.078 &    13.60 &    13.75 & -2.65 & 3\\
J1016+4040      & 2.816 &    13.61 &    14.05 & -2.46 & 5\\
Q1108-077       & 3.607 & $>$14.57 &    14.20 & -1.69 & 6\\
Q1243+307       & 2.526 &    14.00 &    13.69 & -2.77 & 7\\
J1340+1106      & 2.508 & $>$14.42 &    14.43 & -3.12 & 4\\
J1340+1106      & 2.796 & $>$14.64 &    13.40 & -1.65 & 4\\
J1358+6522      & 3.067 &    14.15 &    14.26 & -2.33 & 8\\
J1358+0349      & 2.852 & $>$14.26 &    13.34 & -2.80 & 9\\
J1419+0829      & 3.050 & $>$14.42 &    13.60 & -1.82 & 4\\
J1558-0031      & 2.703 & $>$14.30 &    13.04 & -1.50 & 10\\
J1558+4053      & 2.553 &    14.15 &    14.18 & -2.42 & 5\\
Q2206-199       & 2.076 & $>$14.27 &    13.71 & -2.07 & 5
\enddata
\tablecomments{Column densities are measured in this work. [O/H] is from \citet{2011MNRAS.412.1047C} if included, and from the reference otherwise.}
\tablerefs{(1) \citet{2015ApJ...800...12C} (2) \citet{2010MNRAS.406.1435E} (3) \citet{2001AA...370..426D}
(4) \citet{2011MNRAS.417.1534C} (5) \cite{2008MNRAS.385.2011P} (6) \citet{2008AA...480..349P} (7) \citet{2018ApJ...855..102C} (8) \citet{2014ApJ...781...31C} (9) \citet{2016ApJ...830..148C} (10) \citet{2006ApJ...649L..61O}}
\end{deluxetable}

Finally, we performed an actual blind search for LIAs at $1.7<z<3.4$ using 25 randomly selected quasars (the first 25 in alphanumeric order) from the KODIAQ sample of HIRES spectra \citep{2014ApJ...788..119L,2015AJ....150..111O,2017AJ....154..114O}. Since the \mgii\ doublet and \feii\ multiplet are in the IR (and these objects do not have IR spectra) we searched the \cii\ region directly, starting at the wavelength of the QSO's Lyman-$\alpha$ emission line and ending at the wavelength of \cii\ for the QSO emission redshift. We treated every significant line not readily identified with an interloping multiplet as candidate \cii\ $1334$\AA, and searched the corresponding locations of \cii\ 1036\AA, \oi\ 1302\AA, \siii\ 1260\AA, \siii\ 1526\AA, and \alii\ 1670\AA\ to confirm the identification. This exercise yielded 17 \cii-selected systems, but zero of these were \civ\ non-detections (Figure \ref{fig:highz_ratios} and Table \ref{tab:HIREStable}).  The same sightlines contain 30 additional absorbers detected only in \civ\ (and occasionally \siiv), i.e. not \cii\ or other low-ionization species.

To quantify the (dis)similarity between the absorber population at $z\sim6$ and comparison samples, we perform two-dimensional two-sample Kolmogorov-Smirnov (K-S) tests \citep{1987MNRAS.225..155F}, where the null hypothesis is that both samples have the same (\N{CII}, \N{CIV}) parent distribution. Since available methods for treating censored data are suspect when applied to datasets that consist largely of upper- or lower-limits that fall outside the range of detections (as is the case for, e.g., \civ\ non-detections at $z\gtrsim6$ and saturated absorption in DLAs), we consider two approaches to dealing with limits. First, we assume all non-detections or saturated lines have the values of the corresponding limits. Second, we perform 10000 Monte Carlo iterations in which limits are replaced with values drawn from a uniform distribution between the limit and $\log N=11$ or 16.5 for upper- and lower-limits, respectively, obtaining a distribution of p-values. Since the K-S test statistic measures only the largest separation between cumulative distributions and is insensitive to differences between the tails of the distributions, where both are near 0 or 1, results are likewise insensitive to the exact range of values used to draw substitutes for limits. Since the samples being compared are themselves constructed in disparate fashions and have different limiting sensitivities and possibly different biases, more rigorous statistical testing is unwarranted.

In Table \ref{tab:2DKS} we list the K-S test p-values that result from treating limits as concrete measurements, and the 50th and 95th percentiles of the p-value distributions obtained with the Monte Carlo approach. It is clear that a correspondence between either high-redshift sample and typical $z\sim3$ DLAs is disfavored, while similarity between $z>5.7$ absorbers and MPDLAs is allowed. Unsurprisingly the p-values are larger when the $z>5.7$ sample is limited to only LIAs (those with \cii\ detections). A comparison between LIAs and low redshift DLAs (without metallicity cuts) or LLSs is also unfavorable, because the latter absorbers have associated \civ\ (and often saturated \cii\ in the case of DLAs.)

While these statistical tests do not rule out the possibility of the same distribution for $z>5.7$ and $5<z<5.7$ absorbers, because of the limited sensitivity to the tails of the distributions, it is clear from Figures \ref{fig:highz_ratios} and \ref{fig:xc_highz} that they are different, with more \cii\ non-detections at lower redshift and more \civ\ non-detections at higher redshift. The large p-values from K-S tests comparing the $5<z<5.7$ sample with LLSs may have a similar cause, since both are clustered around the \cii-\civ\ equality line, but have different amounts of limits. The LLS sample is \hi-selected and includes absorbers with no detected metals that would not be detectable at higher redshift. Performing K-S tests between the $5<z<5.7$ and LLS samples with those metal nondetections removed, we find $P_{50}=0.07$ and $P_{95}=0.25$.

\begin{deluxetable}{llll}
\tablecaption{2D K-S Test p-values\label{tab:2DKS}}
\tablehead{\colhead{Comparison} & \colhead{$P_{\rm limits}$} & \colhead{$P_{50}$} & \colhead{$P_{95}$}}
\startdata
$z>5.7$ \& DLAs & 1.2e-6 & 6.5e-10 & 4.7e-9\\
$z>5.7$ \& MPDLAs & 0.060 & 0.015 & 0.039\\
$z>5.7$ \& LLSs & 0.017 & 0.006 & 0.014\\
$z>5.7$ \& HIRES & 7.1e-6 & 3.4e-4 & 4.8e-4\\
$z>5.7$ \& $5<z<5.7$& 0.018 & 0.020 & 0.047\\
LIAs \& MPDLAs & 0.049 & 0.054 & 0.11\\
LIAs \& DLAs & 2.7e-6 & 1.4e-9 & 1.9e-8\\
LIAs \& LLSs & 0.001 & 7.2e-4 & 0.001\\
$5<z<5.7$ \& DLAs &  1.7e-9 & 1.7e-10 & 1.8e-9\\
$5<z<5.7$ \& MPDLAs & 3.9e-6 & 8.9e-5 & 2.9e-4\\
$5<z<5.7$ \& LLSs & 0.068 & 0.192 & 0.39\\
$5<z<5.7$ \& HIRES & 0.002 & 0.006 & 0.010\\
\enddata
\end{deluxetable}

In summary, our search of the literature and archival spectra does not reveal a heretofore unstudied population of $z\sim3$ LIAs with statistically significant incidence. The only plausible analog identified are the MPDLAs, which represent $\lesssim 10\%$ of all DLAs and $<1\%$ of all heavy element absorption systems at $z\sim 3$. LIAs dominate the absorber count at high redshift. If the LIA-MPDLA correspondence is correct it therefore implies that essentially all DLAs are metal-poor at $z>6$.

\subsection{Disappearance of the \civ\ Phase\label{sec:civ}}

\begin{figure*}
\includegraphics[width=\linewidth]{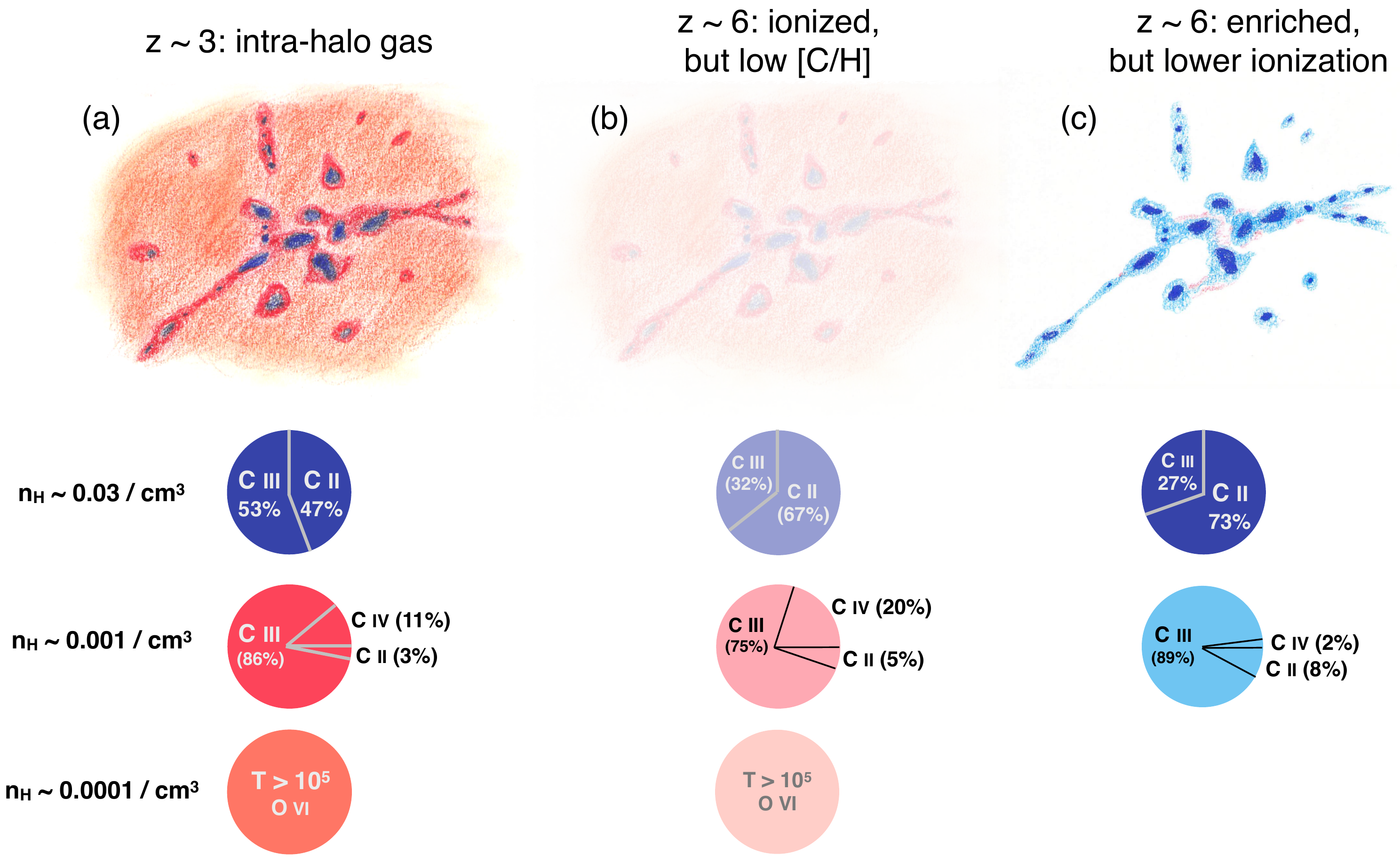}
\caption{Cartoon model of the circumgalactic medium at $z\sim3$, and possible evolutions at $z\sim6$. At low redshift (left), the circumgalactic medium is filled with hot, low density \ovi-absorbing gas (orange), that transitions into higher density regions containing \cii\ (LLSs and DLAs, blue) surrounded by an intermediate density envelope of \cii\ and \civ\ (red). The carbon ionization fractions for these different phases, given in the correspondingly colored charts below, are intended to be illustrative examples from photoionization models. One scenario at high-redshift involves similar ionization states but lower overall metallicity, represented by dimmer colors. The other case illustrated has less drastic metallicity evolution, but different ionization fractions (most likely more \ciii\ at the expense of \civ). The most plausible scenario is a combination of such changes. \label{fig:cartoon}}
\end{figure*}

The disappearance of \civ\ comparable to or greater in strength than \cii\ in LIAs at $z>5.7$ could be caused by low [C/H] values in an otherwise highly-ionized medium, or it could be caused by a change in the ionization conditions of circumgalactic gas that disfavor the triply ionized state. 

In contrast, at $z\sim 3$, the near-universal detection of \civ\ in DLAs, MPDLAs, sub-DLAs and LLSs is explained by invoking a multi-phase model of galactic halos. In this scenario, neutral gas represents a cold and dense precipitate embedded in the hot, low density halo that is more highly ionized \cite[e.g.,][]{2016ApJ...830...87S}. This precipitate may be in the disk, or at radii up to several tenths of the virial radius. The hot phase occupies a larger volume filling factor and may only be visible in \ovi, but \civ\ is mixed into both the hot and cold phases, and may predominantly be in an intermediate density envelope at the interface of these environments, so a single sightline may pierce multiple phases producing absorption overlapping in velocity space \cite[see, e.g.,][]{2015ApJ...802...10C}. \citet{2007A&A...473..791F} find a correlation between metallicity of the neutral phase and \logN{CIV} in DLAs and sub-DLAs at $z=$2--3, suggesting that the different phases may possess a shared enrichment history.

Figure \ref{fig:cartoon} shows a cartoon model of how this paradigm might differ from moderate redshift ($z\sim 3$, left panel), to $z>5.7$ LIAs (center and right). The carbon ionization fractions associated with the various gas phases in the cartoon are drawn from photoionization models described in the following subsection. In one scenario, circumgalactic gas at high redshift retains a similar temperature and ionization parameter, but is sufficiently metal-poor that the \civ\ absorption (red) it produces is too weak to be detected (illustrated by a lower color saturation/opacity). Low-ionization absorption (blue) is detected from the metal-poor neutral precipitate only, but at lower column density. In this case we seek upper limits on the allowed heavy-element abundance.

In the second scenario (right), the disappearance of \civ\ is driven by changes in ionization rather than metallicity, when specific combinations of [C/H], ionization parameter, and spectral shape populate most carbon into \ciii\ state (red), with {\em both} \civ\ and \cii\ (except in DLAs) reduced below our detection thresholds. In this case an enriched and ionized circumgalactic medium could still exist, undetected at $z>6$. A substantial fraction of carbon being in the triply ionized state at $z\sim6$ does not necessitate fine-tuning; \ciii\ is often the dominant ionization state in $z<1$ \citep[e.g.][]{2018ApJ...866...33L}, and photoionization models of $z\sim3$ LLSs \citep{2016ApJ...833..270G}, where \ciii\ is difficult to measure directly, suggest this is still the case. Hence, a relatively modest change in typical ionization conditions could result in the ionization fractions requisite for this scenario.

\subsubsection{Photoionization Modeling}

\begin{figure*}
\plotone{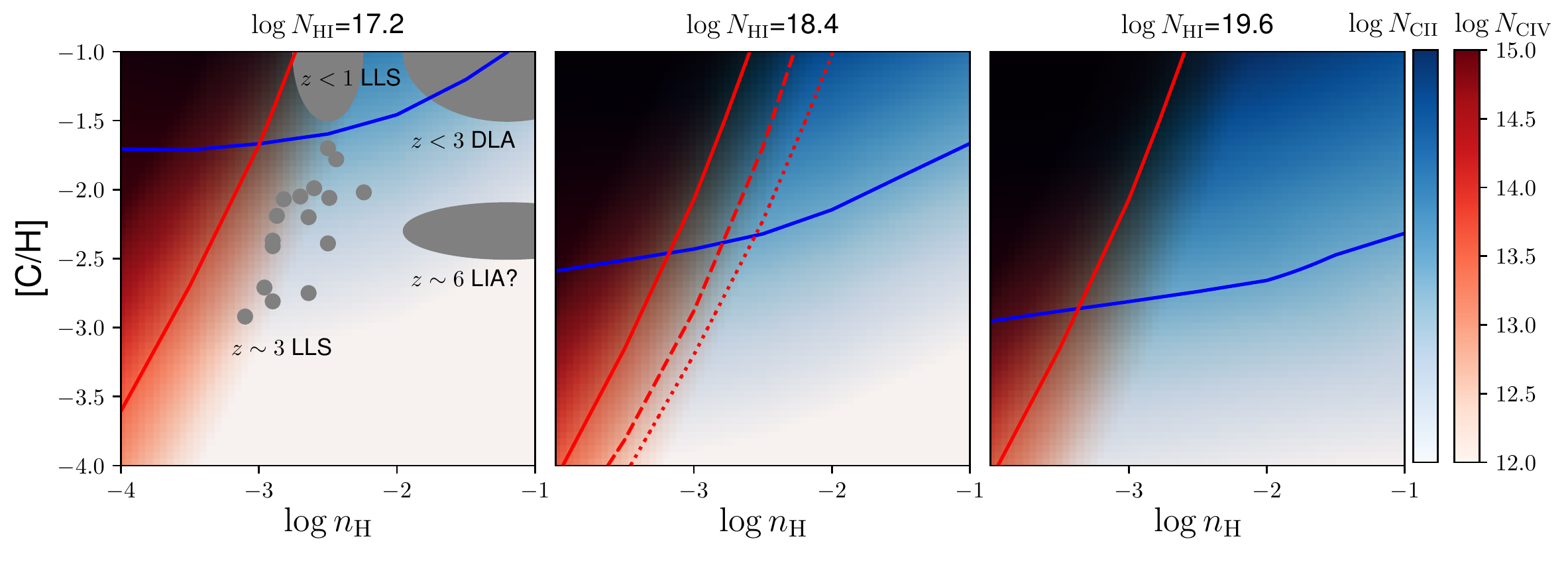}
\caption{\texttt{Cloudy}-calculated column densities of \cii\ (blue) and \civ\ (red) for absorbers of varying \nhi\ at $z=6$. The solid red and blue lines map contours where \logN{CII}=13.5 and \logN{CIV}=13.5 respectively, corresponding to the approximate detection limit in FIRE spectra. The dashed and dotted red curves correspond to \civ\ contours of the $z=3$ UVB and a hardened $z=6$ UVB, respectively.  Absorbers in the lower right of the diagram (shaded white) would be undetected in either species; in the upper left (shaded black) both are detected. Regions shaded red will be detected in \civ\ only, and blue in \cii\ only. Gray ellipses in the left panel annotate approximate locations of various populations, although the DLAs and LIAs may be at larger $n_\mathrm{H}$, and individual LLSs at $z\sim3$ are shown as smaller points.  \label{fig:cloudy_standard}}
\end{figure*}

A more realistic treatment likely involves a combination of these effects, so we use the \texttt{Cloudy} spectral synthesis code (version 13.03, last described by \citealp{2013RMxAA..49..137F}) to explore the parameter space of ionization and abundance.  We model absorbers as plane-parallel, isothermal slabs, illuminated externally by a constant ultraviolet background radiation field (UVB). We use the prescription of \citet{2012ApJ...746..125H}, including sawtooth absorption from \heii\ Lyman-$\alpha$ \citep{2009ApJ...693L.100M} as a baseline model of the metagalactic spectral energy distribution at $z=6$, but also experiment with an ad-hoc prescription to modify the slope. Both the mean spectrum and its spatial variation at 0.7-4.0 Ryd (the energies of interest for carbon ionization balance) are manifestly uncertain once one considers redshifts beyond \heii\ reionization and approaches \hi\ reionization. These models should therefore be viewed as exploratory exercises to understand how circumgalactic matter around primordial galaxies presents observationally---not as precise tools for measuring or correcting individual abundances in the absence of \hi\ measurements.

Figure \ref{fig:cloudy_standard} displays the \texttt{Cloudy} column densities predicted for \cii\ (blue) and \civ\ (red) within a grid of $n_\mathrm{H}$ and [C/H], assuming the fiducial UV background spectrum. The three \hi\ column densities used correspond to the weakest (and most abundant) LLSs at $\log\N{HI}=17.2$,  somewhat larger LLSs at $\logN{HI}=18.4$, and sub-DLAs that may constitute LIAs at $\logN{HI}=19.6$. Solid lines are drawn to illustrate contours of $\logN{CII}=13.5$ and $\logN{CIV}=13.5$, as these values are typical detection limits for both ions in our FIRE spectra. Gray contours on the left-hand plot indicate typical values for several absorber populations at various redshifts (note DLAs and LIAs may have $n_\mathrm{H}$ above the scale on the figure); LLSs at $z=3$ shown in Figure \ref{fig:highz_ratios} are included as smaller gray points. As was described by \cite{2016ApJ...833..270G} for $z\sim3$, a \cii\ non-detection primarily bounds metallicity since it depends only weakly on ionization parameter, whereas non-detection of \civ\ primarily constrains $\log U$ and $n_\mathrm{H}$.

There exists a region of parameter space to the lower right of Figure \ref{fig:cloudy_standard} where absorbers at $z\sim6$ can have LLS or sub-DLA \hi\ column density, and yet remain undetected in either \cii\ or \civ. The precise metallicity limits obtained depend upon the \hi\ column density assumed. For example, at $\logN{HI}\gtrsim 19.6$, \cii\ would be detected for any system with [C/H]$\gtrsim -3$. Because sub-DLAs at this \nhi\ would outnumber observed LIAs (Figure \ref{fig:DLA_lx}), they likely have metallicity of 0.001$Z_\odot$ or smaller (i.e. $\sim 0.5$ dex lower than $z=3-4$ LLSs, and $\sim1$ dex lower than $z\sim 2$ sub-DLAs) or we would likely detect more \cii\ absorbers.

LLSs with $17<\logN{HI}<19$ should be even more numerous, but the \cii-based limit on [C/H] is less stringent. In this parameter space, any gas with [C/H]$<-2$ would also be undetected in \cii\ or \civ, provided $\log n_\mathrm{H} \gtrsim -3$ (using the fiducial $z=6$ UVB, this corresponds to $\log U \lesssim -2$). 

It is striking that LLS at $-3<\rm{[C/H]}<-2$ and $-3<\log n_\mathrm{H}<-2$ would be undetected in metal lines, because this is the exact region of phase space that gives rise to copious \civ\ and \cii\ absorption in the circumgalactic medium at $z\sim 0.5-4.0$ \citep{2015ApJ...812...58C,2016ApJ...833..270G,2016MNRAS.455.4100F,2014ApJ...792....8W,2013ApJ...770..138L}. Why should this region that produces so many strong metal-line systems at lower redshift remain hidden at $z>5.7$?

The most likely explanation recognizes, in addition to presumably lower enrichment at $z\sim6$, the significant softening of the UVB spectrum that occurs with increasing redshift. AGN activity falls off toward higher $z$ \citep[e.g.,][]{2017MNRAS.466.1160M}, and heavy filtering of source radiation from absorption in the \heii\ continuum ($>4$ Ryd), \heii\ Lyman-$\alpha$ forest ($3-4$ Ryd), \hi\ continuum ($>1$ Ryd) and \hi\ Lyman-$\alpha$ forest ($0.7-1$ Ryd) drastically reduces the number of $>3.5$ Ryd photons available to ionize \ciii\ into \civ\ at fixed $n_\mathrm{H}$.

\begin{figure}[t]
\plotone{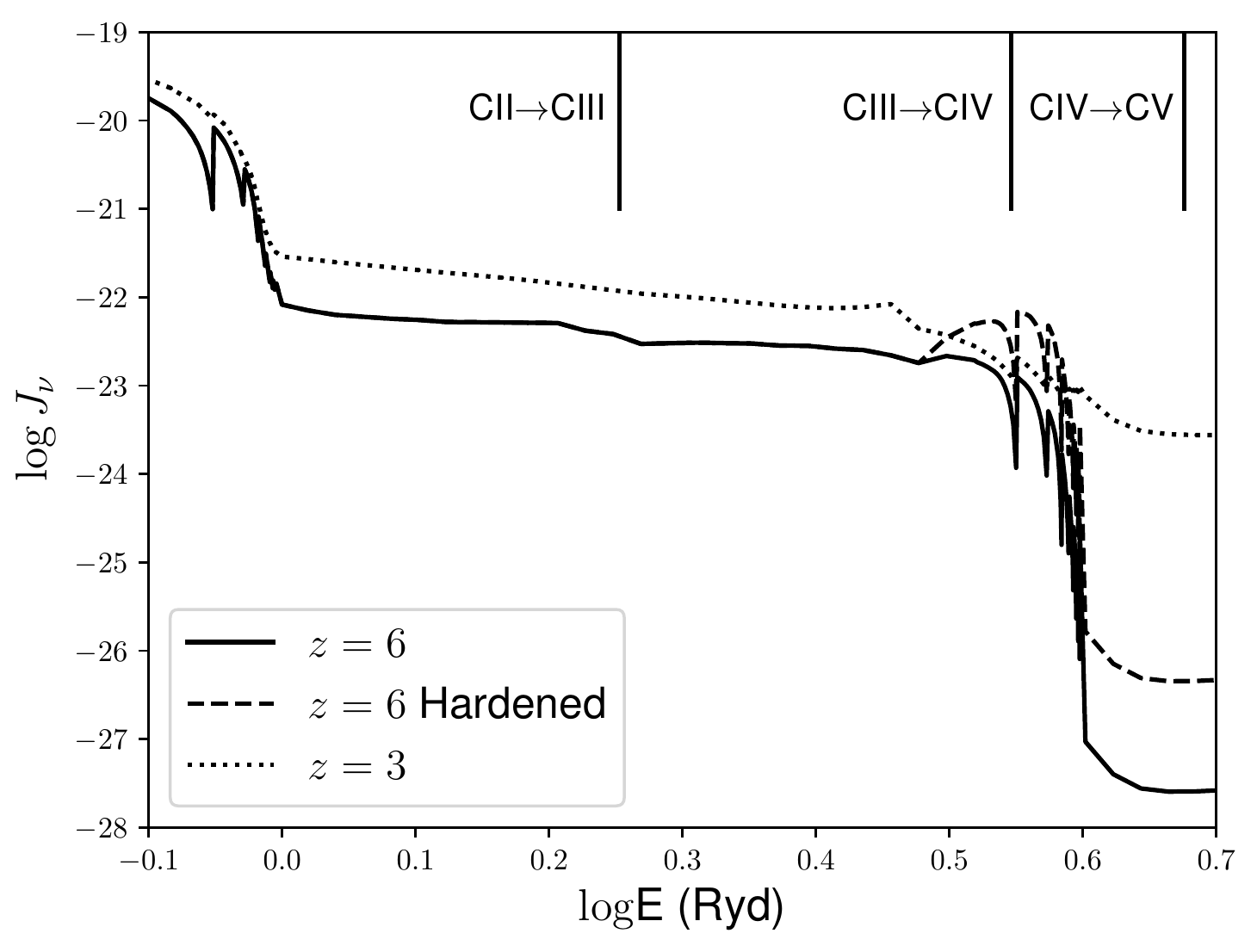}
\caption{The UV background at $z=6$ (solid), $z=3$ (dotted), and $z=6$ hardened via Equation \ref{eq:uvb}. The ionization energies of carbon ions are indicated. The $z=3$ and hardened spectra both result in larger \civ\ column densities for LLSs, as demonstrated in Figure \ref{fig:cloudy_standard}.}\label{fig:UVB}
\end{figure}

Put another way, the harder UVB spectrum at $z\lesssim 4$ shifts the red contour rightward to the point where \civ\ becomes detectable in LLSs at $-3<\log n_\mathrm{H} < -2$, over a wide range in [C/H]. We demonstrate this in Figure \ref{fig:cloudy_standard} with red lines corresponding to the $\log\N{CIV}=13.5$ contours for the $z=3$ UVB (dotted) and for a hardened version of the $z=6$ UVB (dashed) at $\log\N{HI}=18.4$. The $z=3$ \citep[also from][]{2012ApJ...746..125H}, $z=6$, and hardened $z=6$ UVB spectra are shown in Figure \ref{fig:UVB}. For the hardened spectrum, we increase the UV flux by a factor of $\sim18$ beyond the \heii\ ionization edge, and smoothly transition from the nominal spectrum across the \heii\ Lyman-$\alpha$ forest, a range which also covers the \ciii\ ionization energy:
\begin{equation}
J_\nu^{'}(E)=
\begin{cases}J_\nu(E)\left(\frac{E}{3\rm{~Ryd}}\right)^{10} & 3<E<4\rm{~Ryd}\\
J_\nu(E)\left(\frac{4\rm{~Ryd}}{3\rm{~Ryd}}\right)^{10} & E>4\rm{~Ryd}
\end{cases}\label{eq:uvb}
\end{equation}
 Both this simple modification at $z=6$ and the nominal $z=3$ UVB produce \civ\ contours shifted relative to the nominal $z=6$ UVB such that lower metallicity systems of the same density produce observable \civ\ absorption. Given the rise in $f_{\rm HI}(N,X)$ toward lower \nhi\ across the LLS regime, this suggests an explanation for how \civ\ absorbers evolve from being a small minority of all metal-line systems at $z>5.7$ to the dominant population at lower redshift.

Circumgalactic gas at $z\sim 6$ has either not yet been enriched above $\sim$1/1000 Solar, or else it has 1/1000--1/300 Solar metallicity but a weak radiation field at high energies, which sequesters most carbon into the unobservable \ciii\ state. The gradual hardening of the UV background leads to a concomitant rise in \ciii$\rightarrow$\civ\ ionization, revealing metal absorption in this region of [C/H]-$n_\mathrm{H}$ parameter space.  As \civ\ absorption exceeds the detection threshold for LLSs of progressively lower \nhi, $\dNdXt$ for \civ\ increases accordingly.

As a final detail, we note that our \texttt{Cloudy} modeling likely \textit{underpredicts} \N{CIV} (and possibly \N{CIII}). The high-ionization absorption largely arises from a comparatively low-density envelope that surrounds the denser neutral gas (see Figure \ref{fig:cartoon}), but our models treat absorbers as uniform density. The lower density of the envelope suggests a larger fraction of \N{CIII} and \N{CIV}. In effect, accounting for this underprediction would disfavor LLSs even more. LLSs require higher metallicity than DLAs to match a given \N{CII} observation (e.g., $\log\N{HI}=17.2$ requires [C/H]$=-1.5$ to produce $\log\N{CII}=13.5$), but the underprediction of \civ\ means that \textit{lower} abundances are required in order for \civ\ to go unobserved.

That is, to adequately predict the absorption signature a model needs to account for the non-uniform density, and doing so will increase the absorption from higher-ionization species without much effect on \cii. In effect, this underprediction of \N{CIV} means the above discussion is less constraining on LLSs than a more sophisticated approach: the red shading in Figure \ref{fig:cloudy_standard} should be shifted to the right.

\subsubsection{Complimentary Means of Changing \N{CII}/\N{CIV}}

As galaxies evolve, changes in characteristics such as clumpiness and AGN activity may play a role in shaping the CGM, in conjunction with the changing metallicity and ionization fractions.  We briefly consider several features that could impact CGM observations, but note that an in-depth analysis requires modeling and/or simulations beyond the scope of this work.

One could reasonably ask if, in addition to changing ionization fractions, weaker feedback at high redshift leads to a smaller cross-section of the diffuse CGM that gives rise to high-ionization absorption. Assuming a simple model where \cii\ is embedded in a diffuse halo of \civ\ gas, the changing \cii-to-\civ\ ratio of Figure \ref{fig:xc_highz} would then imply that the radius of the \civ-bearing CGM region must shrink by a factor of $\sim10-100$ between $z=5$ and 6, such that this gas is coincident with the stellar component of the galaxy rather than a surrounding halo. Galaxy formation simulations do show that galaxies are still assembling their hot gaseous halos at $z=6$ \citep[e.g.,][]{2018MNRAS.473..538C}, but halos are certainly not nonexistent, and halo growth continues to lower redshifts such that the \cii-to-\civ\ ratio should continue to decrease if halo growth has a strong effect on absorption properties. Comparing the $5.0<z<5.7$ and $2.0<z<3.5$ absorbers in Figure, \ref{fig:highz_ratios} seems to suggest this is not the case, although both samples largely consist of \cii\ nondetections so this comparison is limited.

Reduced mixing with the outer CGM may be another avenue to explain decreasing \civ\ at high redshift. If feedback does not inject enough energy to drive metals produced in star-forming regions well beyond the ISM, then gas that would bear \civ\ would be lower metallicity than that with \cii. Hence, a plausible complement to changing ionization fractions in explaining the observations is a lack of enrichment of the outer halos. As the photoionization models we implement assume uniform metallicity, the effect of reduced mixing would allow for softer UV background spectra and/or higher gas densities that increase the \civ\ fraction in less enriched regions while having less impact on \cii. While a full discussion of this requires simulations to explore the role and effect of feedback on nascent galaxies, we note that, similar to changing cross-sections, metallicities in outer halos would have to be $\sim0.1-0.01$ times that of regions producing \cii\ absorption.

\subsection{\cii\ Mass Density ($\Omega_\mathrm{CII}$) Estimate}

Using our low-ionization measurements, we may estimate the universal mass density of \cii\ ions, expressed in the common form of its contribution to closure density, $\Omega_\mathrm{CII}$. This formulation, which represents the first moment of the \cii\ column density distribution function, can be calculated from the discrete measurements of \N{CII} \citep{1996MNRAS.283L..79S}:
\begin{eqnarray}
\Omega_\mathrm{CII}&=&\frac{H_0m_c}{c\rho_\mathrm{crit}}\frac{\sum_i N_{\mathrm{CII},i}}{\Delta X}\\
\left(\frac{\sigma_{\Omega}}{\Omega_{\mathrm{CII}}}\right)^2&=&\frac{\sum_i N_{\mathrm{CII},i}^2}{\left(\sum_i N_{\mathrm{CII},i}\right)^2}.
\end{eqnarray}
where $H_0$ is the Hubble parameter, $m_c$ is the mass of a carbon atom, $\rho_c$ is the critical density, $c$ is the speed of light, and $\Delta X$ is the total absorption pathlength surveyed (see Section \ref{sec:incidence}). Since the sum is dominated by systems of high \N{CII}, it is conveniently robust with respect to incompleteness at low column density.  

We calculate $\Omega_\mathrm{CII}/10^{-8}=0.70\pm0.20$ at $z>5.7$, using all detections of \cii\ in our FIRE sample spectra. If we include pathlength where \mgii\ is detected in lieu of \cii\ and convert using Equation 1, the resultant estimate is ($0.67\pm0.18$). For both estimates, non-detections were added into the sum at the value of their $3\sigma$ upper limits.  Again, because the integral is dominated by the high column density tail, the convergence does not depend strongly on how these limits are treated. For example, if we added the non-detections at identically \N{CII}$=0$, then $\Omega_\mathrm{CII}$ only decreases by $\sim 5\%$.

Saturated \cii\ lines are potentially a larger source of inaccuracy. For these we assign \N{CII} to the value of its measured lower limit, which may be well below the actual value. If all of these systems instead had $\logN{CII}=15.0$, then $\Omega_\mathrm{CII}$ could increase by a factor of 2-3.

Our estimates for $\Omega_\mathrm{CII}$ are broadly consistent with those of \citet{2011ApJ...735...93B}, who measure $\Omega_\mathrm{CII}/10^{-8}\ge0.9$, about 25\% higher than our measurement but still in statistical agreement. They are also comparable to the value of $\Omega_\mathrm{CIV}$ at $z\sim5.7$ \citep{2011ApJ...743...21S,2013MNRAS.435.1198D}, consistent with the general notion that \civ\ is not globally dominant to \cii\ at high redshift.

The \cii\ density at later epochs ($z<5.7$) is $\sim 30\%$ smaller (but again statistically consistent with no evolution) at $\Omega_\mathrm{CII}/10^{-8}=0.44\pm0.24$.

Because the mass density of triply ionized carbon declines by nearly an order of magnitude from $z=1$ to $z=5$ \citep{2010MNRAS.401.2715D,2013ApJ...763...37C}, and the mass density of singly ionized carbon remains fairly flat over the same range \citep[assuming it evolves similarly to the \mgii\ incidence rate,][]{2017ApJ...850..188C}, it follows that the sum $\Omega_{\mathrm CII}+\Omega_{\mathrm CIV}$ cannot remain constant and must decline toward higher $z$. If the decline in \civ\ is driven by ionization, then the picture must be more subtle than a zero-sum conversion of \civ\ into \cii\, because we do not observe nearly a large enough increase in $\Omega_\mathrm{CII}$ to offset the lost \civ.

It is tempting to interpret the declining value of $\Omega_\mathrm{CII}+\Omega_\mathrm{CIV}$ toward high redshift as a direct signature of chemical enrichment, since the two ions together trace both low- and high-ionization gas and therefore account for a larger fraction of all circumgalactic carbon atoms.  However this simple picture does not recognize that the preponderance of carbon atoms are quite possibly in the \ciii\ state, which cannot be observed because \ciii\ $\lambda$977\AA ~is in the Lyman$-\alpha$ forest.

Examination of these trends together with Figure \ref{fig:cloudy_standard} suggests that DLAs dominate the budget of $\Omega_\mathrm{CII}$ at all redshifts, and that any redshift trends in this ion may be attributed to chemical enrichment (since Figure \ref{fig:cloudy_standard} shows that \cii\ is less sensitive to $n_\mathrm{H}$ than \civ). Indeed low abundances are required to avoid saturation in the \cii\ profiles of LIAs as observed, if they have \hi\ column densities in range of DLAs.

In contrast the \civ\ frequency is affected by both abundances and ionization.  Our observations do not distinguish between a scenario where (a) the non-detection of a highly ionized circumgalactic medium results from a very low heavy-element abundance, or (b) a softening of the UV spectrum at $z>5$ favors the transfer of triply ionized circumgalactic \civ\ into doubly ionized \ciii, obscuring early enriched matter from view. This needs to happen in a way that does not also increase the concentration of \cii\, but it seems this may be possible, given the relative insensitivity of \cii\ in DLAs to ionization (the sensitivity of \ciii\ to ionization is comparable to that of \civ).

The cosmological hydrodynamical simulations of \citet{2015MNRAS.447.2526F} predict a mass density at $z\sim6$ of $\Omega_\mathrm{CII}/10^{-8}\sim0.3$, about a factor of two less than we estimate. They also show that both absorber incidence rate and mass density of \cii\ evolve much more slowly than those of \civ\ at high-redshift; for \cii\ both quantities increase by a factor of a few between $z=9$ to $z=6$ while those of \civ\ grow by two orders of magnitude. They additionally demonstrate the presence of large scale fluctuations in the UVB at early times, and \citet{2016MNRAS.459.2299F} show that \cii\ is found across a broad range of UVB intensities, while \civ\ is largely restricted to the higher end of UVB intensities. This suggests the differences in the evolution of \cii\ and \civ\ are ionization-driven, in the sense that quantifications of \civ\ are more sensitive to ionization conditions assuming similar enrichment of regions with different ionization states.

\section{Summary}

We have presented concurrent analysis of low- and high-ionization species in 69 $z>5$ absorption systems, including several spectroscopically resolved low-ionization systems with accompanying high SNR coverage of more highly ionized species. Comparing \cii\ and \civ\ column densities of individual absorbers, we find that high-redshift absorption systems are typically dominated by the low-ionization phase, increasingly so at $z\gtrsim5.7$, opposite the situation at lower redshifts.  \textit{Half} of the absorbers at $z\ge5.7$ have no \civ\ detections, and six have relatively low limits on or measurements of \N{CIV} ($\lesssim13.0$) such that they must have \textit{at least} ten times as much absorption from the low-ionization phase. At $5.0<z<5.7$, only 20\% of absorbers exhibit only the low-ionization phase, and about half show \civ\ only.
 
These absorbers, which we dub LIAs, seemingly have no common $z\sim3$ analog with strong \cii\ and very weak or undetected \civ. LLSs have comparable \cii\ but have a similar amount of coincident \civ, and DLAs generally have strong (often saturated) \cii\ and \civ. The most promising candidates are metal-poor DLAs ([O/H]$\lesssim-2$), that represent $\lesssim10\%$ of all DLAs at $z\sim3$.

Following statistical arguments based on incidence rates as a function of \N{HI}, we conclude that LIAs (with $\dNdXt\sim0.17$) are consistent with a source population with \hi\ in the range of DLAs and sub-DLAs. Moreover, if we assume they instead arise from weaker \hi\ absorbers such as LLSs, then we would expect far more LIAs than seen. This implicitly requires that the majority of DLAs and sub-DLAs at $z\gtrsim6$ have metallicities of [C/H]$\lesssim-2.5$. In order to go undetected, LLSs must have [C/H]$<-2$, as is already typical at $z\sim3$.

Resolved spectra of LIAs at $z\gtrsim6$ exhibit narrow kinematics, generally contained within an envelope of $\lesssim100\kms$, less broadened than \civ\ profiles at $z<5.7$. The temperatures of these narrow absorption systems ($T\lesssim10^4$ K, derived from Voigt profile fitting) are comparable to that of low-ionization species in the low redshift circumgalactic medium; there are also absorbers or components with low temperatures of $T<200$ K. These resolved profiles are accompanied by one of our highest quality IR spectra at these redshifts, providing the largest \cii-to-\civ\ ratios.

Our results suggest a significant change to circumgalactic gas presumed to give rise to these absorption systems, with a combination of lower chemical abundances and a softer ionizing UV background at $z\sim6$ in comparison to $z=3$, but cannot distinguish the relative contributions of these factors. Simulations suggest that improved observations of numerous low- and high-ionization species may yield sufficient discriminatory data \citep{2016MNRAS.459.2299F,2018MNRAS.475.4717D}. As more high-redshift quasars are observed and more high-SNR spectra are obtained \citep[e.g.,][]{2017MNRAS.470.1919B}, increased absorption pathlength and sensitivity will provide more detail on the nature of LIAs and the apparent disappearance of absorption from the high-ionization phase. Additionally, identification of high-redshift galaxies associated with absorption systems with the upcoming James Webb Space Telescope may provide critical information on the environments in which LIAs exist.

\acknowledgements This paper includes data gathered with the 6.5 meter Magellan Telescopes located at Las Campanas Observatory, Chile. Some of the data presented herein were obtained at the W. M. Keck Observatory, which is operated as a scientific partnership among the California Institute of Technology, the University of California and the National Aeronautics and Space Administration. The Observatory was made possible by the generous financial support of the W. M. Keck Foundation. The authors wish to recognize and acknowledge the very significant cultural role and reverence that the summit of Maunakea has always had within the indigenous Hawaiian community.  We are most fortunate to have the opportunity to conduct observations from this mountain. This research has made use of the Keck Observatory Archive (KOA), which is operated by the W. M. Keck Observatory and the NASA Exoplanet Science Institute (NExScI), under contract with the National Aeronautics and Space Administration. Some of the data presented in this work were obtained from the Keck Observatory Database of Ionized Absorbers toward QSOs (KODIAQ), which was funded through NASA ADAP grant NNX10AE84G. This research has made use of the services of the ESO Science Archive Facility. RB was supported by NASA through Hubble Fellowship grant \#51354 awarded by the Space Telescope Science Institute, which is operated by the Association of Universities for Research in Astronomy, Inc., for NASA, under contract NAS 5-26555. KLC acknowledges support from NSF grant AST-1615296 and appreciates the observational support of A. Hedglen, a University of Hawai`i at Hilo undergraduate at the time.

\bibliography{hiz}
\appendix

Two additional data tables are provided in this appendix. Table \ref{tab:J0100abs2} details the measurements of individual Voigt profile components of absorbers along the line of sight to QSO J0100+2802, described in Section \ref{sec:J0100}. Table \ref{tab:HIREStable} lists the \cii\ and \civ\ column densities for absorption systems identified in a survey of 25 archival HIRES quasar spectra, described in Section \ref{sec:analogs}.

\setcounter{table}{0}
\renewcommand{\thetable}{A\arabic{table}}
\begin{deluxetable}{rcccccccccc}[h]
\tablecaption{J0100+2802 Absorption System Voigt Profile Fit Components\label{tab:J0100abs2}}
\tablehead{
\colhead{$z_{\text{abs}}$} & \colhead{$\log\N{CII}$} & \colhead{$\log\N{SiII}$} &\colhead{$\log\N{OI}$} & \colhead{$\log\N{SiIV}$}  &\colhead{$\log\N{CIV}$} & \colhead{\bD{C}} & \colhead{\bD{Si}} &\colhead{\bD{O}} &\colhead{\bD{turb}} &\colhead{$\log T$}}
\startdata
6.1873 & $<11.79$ & \xpm{11.59}{0.06}{0.07} & $<12.55$ & \phn{}--- & & &  \xpm{2.1}{1.0}{0.8}   \\\hline
6.1431 & $<12.96$ & \xpm{12.38}{0.10}{0.10}  & \xpm{13.80}{0.07}{0.07} & $<12.36$ & \phn{}--- & \xpm{6.4}{2.2}{1.0}    & \xpm{6.2}{0.8}{0.8}& \xpm{6.4}{1.6}{1.0}  & \xpm{5.7}{0.8}{0.6} & \xpm{1.9}{2.6}{1.6}\\
 .1434 & \xpm{13.79}{0.03}{0.04} & \xpm{12.81}{0.05}{0.04} & \xpm{14.37}{0.03}{0.03} & $<12.55$ & \phn{}--- & \xpm{18.7}{1.0}{0.9} & \xpm{15.2}{0.9}{1.1}& \xpm{17.3}{0.6}{0.6}& \xpm{12.2}{1.9}{4.6}  &\xpm{5.1}{0.2}{0.2}\\
 .1438 & \xpm{13.86}{0.04}{0.03} & \xpm{13.20}{0.04}{0.03}  & \xpm{14.37}{0.04}{0.04} & $<12.38$ & \phn{}--- & \xpm{7.9}{0.4}{0.4}    & \xpm{5.8}{0.2}{0.2}& \xpm{7.1}{0.3}{0.3}  & \xpm{3.3}{0.6}{0.4} & \xpm{4.6}{0.1}{0.1}\\ \hline
6.1115 & \xpm{13.73}{0.04}{0.04}  & \xpm{12.78}{0.02}{0.02} & \xpm{14.28}{0.04}{0.03} & $<11.86$ & \phn{}--- & \xpm{5.0}{0.3}{0.2} &  \xpm{5.0}{0.2}{0.2} & \xpm{5.0}{0.2}{0.2} &\xpm{4.9}{0.2}{0.3} & \xpm{2.2}{1.4}{1.5}\\
 .1118 & \xpm{13.40}{0.04}{0.04} & \xpm{12.42}{0.02}{0.01}     & \xpm{13.90}{0.02}{0.02} & $<11.73$ & \phn{}--- & \xpm{5.2}{0.3}{0.2} &  \xpm{5.1}{0.3}{0.2} & \xpm{5.1}{0.3}{0.2} &\xpm{5.1}{0.3}{0.3} & \xpm{2.1}{1.2}{1.4}\\ \hline
5.7973 & \xpm{14.04}{0.22}{0.05} & \xpm{13.49}{0.01}{0.02} & \phn{}--- & $<11.75$ & \phn{}--- & \xpm{5.9}{0.3}{1.0} & \xpm{4.3}{0.5}{0.2} & \phn{}--- & \xpm{2.0}{2.6}{0.8} & \xpm{4.4}{0.1}{2.0}\\
 .7978 & \xpm{13.65}{0.02}{0.01} & \xpm{13.08}{0.03}{0.03} & \phn{}--- & $<12.03$ & \phn{}--- & \xpm{8.1}{0.3}{0.3} & \xpm{7.9}{0.3}{0.3} & \phn{}--- & \xpm{2.0}{2.6}{0.8} & \xpm{4.4}{0.1}{2.0}\\ \hline
5.3381 & \phn{}--- &$<12.77$ & \phn{}--- & \xpm{12.83}{0.02}{0.02} & \xpm{13.73}{0.02}{0.02} & \xpm{16.4}{0.6}{0.6} & \xpm{10.9}{0.4}{0.4} & \phn{}--- & \xpm{1.8}{1.4}{0.6} & \xpm{5.3}{0.04}{0.04}\\
 .3386 & \phn{}--- &$<12.66$ & \phn{}--- & \xpm{12.70}{0.03}{0.03} & \xpm{13.06}{0.08}{0.09} & \xpm{14.0}{0.9}{1.1} & \xpm{13.9}{0.7}{1.0} & \phn{}--- & \xpm{13.6}{0.9}{1.1} & \xpm{2.2}{2.2}{1.6}\\
 .3393 & \phn{}--- &$<12.54$ & \phn{}--- & \xpm{12.52}{0.04}{0.04} & \xpm{13.38}{0.06}{0.06} & \xpm{23.8}{2.8}{2.8} & \xpm{16.6}{1.7}{1.7} & \phn{}--- & \xpm{6.9}{3.4}{3.2} & \xpm{5.6}{0.1}{0.2}\\ \hline
5.1080\tablenotemark{a} & --- & --- & --- & --- & \xpm{13.64}{0.06}{0.17} & \xpm{23.9}{1.8}{6.2} & --- & --- & --- & ---\\
 .1081 & --- & --- & --- & --- & \xpm{13.53}{0.15}{0.13} & \xpm{59.5}{0.4}{4.7} & --- & --- & --- & ---\\
 .1091 & --- & --- & --- & --- & \xpm{13.92}{0.02}{0.02} & \xpm{23.0}{3.2}{1.8} & --- & --- & --- & ---\\
 .1100 & --- & --- & --- & --- & \xpm{13.55}{0.02}{0.02} & \xpm{14.7}{0.9}{0.67} & --- & --- & --- & ---\\
 .1108 & --- & --- & --- & --- & \xpm{13.55}{0.03}{0.03} & \xpm{7.3}{0.6}{0.6} & --- & --- & --- & ---\\
 .1115 & --- & --- & --- & --- & \xpm{13.19}{0.04}{0.04} & \xpm{11.0}{3.5}{2.3} & --- & --- & --- & ---\\
 .1121 & --- & --- & --- & --- & \xpm{12.92}{0.07}{0.14} & \xpm{6.2}{1.7}{3.6} & --- & --- & --- & ---\\
 .1133 & --- & --- & --- & --- & \xpm{13.08}{0.04}{0.05} & \xpm{5.2}{2.0}{1.2} & --- & --- & --- & ---\\
 .1137 & --- & --- & --- & --- & \xpm{13.49}{0.02}{0.03} & \xpm{8.6}{1.0}{1.4} & --- & --- & --- & ---\\
 .1140 & --- & --- & --- & --- & \xpm{12.85}{0.11}{0.13} & \xpm{9.6}{1.9}{1.9} & --- & --- & --- & ---\\
& & (\N{Fe II}) & & & & & (\bD{Fe II})\\
5.1075 & --- & \xpm{14.14}{0.03}{0.03} & \xpm{13.77}{0.04}{0.04} & --- & --- & --- & \xpm{20.7}{1.6}{1.4} & \xpm{20.4}{1.7}{1.4} & \xpm{20.3}{1.7}{1.6} & \xpm{3.0}{1.8}{2.2} \\
 .1079 & --- & \xpm{13.03}{0.20}{0.27} & $<12.71$               & --- & --- & --- & \xpm{3.7}{11.1}{1.8} & \xpm{3.6}{11.0}{1.8} & \xpm{3.4}{7.8}{1.8} & \xpm{2.4}{1.7}{1.7} \\
 .1083 & --- & \xpm{13.87}{0.04}{0.04} & \xpm{13.52}{0.05}{0.06} & --- & --- & --- & \xpm{18.1}{1.3}{1.2} & \xpm{17.9}{1.4}{1.2} & \xpm{17.8}{1.4}{1.4} & \xpm{3.0}{1.6}{1.9}\\\hline
4.8748 & \phn{}--- &$<12.20$ & \phn{}--- & ---& \xpm{12.86}{0.02}{0.02} & \xpm{9.06}{0.76}{0.68} & --- & --- & --- & ---\\
 .8754 & \phn{}--- &$<12.25$ & \phn{}--- & ---& \xpm{12.81}{0.03}{0.03} & \xpm{11.12}{1.05}{0.97} & --- & --- & --- & ---\\
\enddata
\tablenotetext{a}{Low- and high-ionization species are fit independently for this absorber.}
\end{deluxetable}

\clearpage

\begin{deluxetable}{llll}
\tabletypesize{\footnotesize}
\tablecaption{Low Redshift HIRES Absorbers\label{tab:HIREStable}}
\tablehead{\colhead{Quasar Name} & \colhead{z} & \colhead{$\logN{CII}$} & \colhead{$\logN{CIV}$}}
\startdata
J004530$-$261709 & 3.407 & $<$13.2 & 13.7     \\
J020346$+$113445 & 3.387 & $>$15.1 & 14.0     \\
J004530$-$261709 & 3.372 & $<$13.1 & 13.4     \\
J004530$-$261709 & 3.255 & $>$14.0 & 14.2     \\
J004530$-$261709 & 3.129 & $<$13.0 & 13.4     \\
J004434$-$261121 & 3.103 & $<$13.6 & 13.7     \\
J014516$-$094517 & 2.736 & $<$12.9 & 13.1     \\
J020455$+$364917 & 2.690 & 13.5    & 13.9     \\
J010311$+$131617 & 2.664 & $<$12.6 & 12.7     \\
J012156$+$144823 & 2.664 & $>$14.8 & 14.5     \\
J000150$-$015940 & 2.636 & $<$13.3 & 13.1     \\
J004351$-$265128 & 2.599 & 14.3    & $>$15.0  \\
J010311$+$131617 & 2.552 & 12.1    & 13.3     \\
J002127$-$020333 & 2.537 & $<$13.3 & 13.8     \\
J010806$+$163550 & 2.536 & 14.1    & 13.2     \\
J004358$-$255115 & 2.401 & 13.2    & 14.2     \\
J003501$-$091817 & 2.376 & $<$13.1 & 13.3     \\
J002127$-$020333 & 2.359 & $<$13.2 & 13.5     \\
J010806$+$163550 & 2.356 & $<$12.6 & 13.2     \\
J004351$-$265128 & 2.343 & $>$15.2 & $>$15.2  \\
J015234$+$335033 & 2.337 & $<$13.2 & 13.4     \\
J005700$+$143737 & 2.333 & $<$13.2 & 13.1     \\
J002830$-$281704 & 2.315 & $<$13.6 & 13.9     \\
J005814$+$011530 & 2.288 & $<$13.3 & 13.7     \\
J000931$+$021707 & 2.278 & $<$13.3 & 13.9     \\
J004358$-$255115 & 2.249 & $<$13.2 & 13.8     \\
J005814$+$011530 & 2.236 & $<$13.3 & 13.6     \\
J003501$-$091817 & 2.221 & $<$13.0 & 13.9     \\
J015234$+$335033 & 2.204 & 13.2    & 14.2     \\
J005202$+$010129 & 2.192 & 13.4    & 14.3     \\
J022839$-$101110 & 2.171 & $<$13.0 & 13.4     \\
J002952$+$020606 & 2.168 & $<$13.4 & 13.5     \\
J015234$+$335033 & 2.141 & $>$14.6 & 13.9     \\
J000931$+$021707 & 2.130 & $<$13.2 & 13.1     \\
J002952$+$020606 & 2.121 & 14.1    & $>$14.8  \\
J001602$-$001224 & 2.029 & $>$14.4 & $>$14.7  \\
J015227$-$200107 & 2.010 & $<$13.5 & 13.9     \\
J005202$+$010129 & 1.962 & $<$13.1 & 13.5     \\
J023145$+$132254 & 1.958 & 13.3    & 13.9     \\
J015227$-$200107 & 1.929 & $<$13.3 & 13.7     \\
J023145$+$132254 & 1.903 & 13.2    & 14.4     \\
J023145$+$132254 & 1.862 & $>$14.7 & $>$14.4  \\
J023145$+$132254 & 1.841 & $<$13.2 & 13.0     \\
J000520$+$052410 & 1.802 & $<$13.0 & 12.9     \\
J000520$+$052410 & 1.778 & $<$13.1 & 12.9     \\
J000520$+$052410 & 1.745 & 12.7    & 14.2     \\
J012227$-$042127 & 1.740 & $<$13.2 & 13.8
\enddata
\end{deluxetable}

\end{document}